\documentclass[11pt]{article}
\usepackage{amsmath,amssymb, mcite}
\usepackage{rotating}
\usepackage{caption}

\newcommand{\fr}{\mathfrak{f}_{\scriptscriptstyle{F \hspace{-0.8mm} R}}}

\newcommand{\tA}{{\tt A}}
\newcommand{\tM}{{\tt M}}
\newcommand{\tN}{{\tt N}}
\newcommand{\tP}{{\tt P}}
\newcommand{\tQ}{{\tt Q}}
\newcommand{\tR}{{\tt R}}
\newcommand{\tS}{{\tt S}}
\newcommand{\tT}{{\tt T}}
\newcommand{\tU}{{\tt U}}
\newcommand{\tV}{{\tt V}}
\newcommand{\tW}{{\tt W}}

\newcommand{\cV}{{\cal V}}
\newcommand{\cM}{{\cal M}}
\newcommand{\cN}{{\cal N}}
\newcommand{\cA}{{\cal A}}
\newcommand{\cB}{{\cal B}}
\newcommand{\cD}{{\cal D}}
\newcommand{\cQ}{{\cal Q}}
\newcommand{\cP}{{\cal P}}
\newcommand{\cR}{{\cal R}}

\newcommand{\cU}{{\cal U}}

\begin{document}

\begin{titlepage}
\hfill DAMTP-2013-72

\hfill  AEI-2013-270 \hspace*{3.2mm}
\vspace{2.5cm}
\begin{center}

{{\LARGE  \bf The embedding tensor of Scherk-Schwarz flux\\[3mm] compactifications from eleven 
dimensions }} \\

\vskip 1.5cm {Hadi Godazgar$^{\star}$ 
, Mahdi Godazgar$^{\dagger}$ 
and Hermann Nicolai$^{\ddagger}$}
\\
{\vskip 0.5cm
$^{\star \dagger}$DAMTP, Centre for Mathematical Sciences,\\
University of Cambridge,\\
Wilberforce Road, Cambridge, \\ CB3 0WA, UK\\
\vskip 0.5cm
$^{\ddagger}$Max-Planck-Institut f\"{u}r Gravitationsphysik, \\
Albert-Einstein-Institut,\\
Am M\"{u}hlenberg 1, D-14476 Potsdam, Germany}
{\vskip 0.35cm
$^{\star}$H.M.Godazgar@damtp.cam.ac.uk, $^{\dagger}$M.M.Godazgar@damtp.cam.ac.uk, 
$^{\ddagger}$Hermann.Nicolai@aei.mpg.de}
\end{center}

\vskip 0.35cm

\begin{center}
\today
\end{center}

\noindent

\vskip 1.2cm

\begin{abstract}
\noindent We study the Scherk-Schwarz reduction of $D=11$ supergravity with background fluxes 
in the context of a recently developed framework pertaining to $D=11$ supergravity.  We 
derive the embedding tensor of the associated four-dimensional maximal gauged theories 
{\em directly} from eleven dimensions by exploiting the generalised vielbein postulates, and
by analysing the couplings of the full set of 56 electric and magnetic gauge fields to the generalised vielbeine. The treatment presented here will apply more generally to other reductions of $D=11$ supergravity to maximal gauged theories in four dimensions.
\end{abstract}

\end{titlepage}

\section{Introduction}

Recently, a reformulation \cite{GGN13} of $D=11$ supergravity \cite{CJS} that 
emphasises the exceptional E$_{7(7)}$ duality symmetry \cite{cremmerjulia} 
and is based on the SU(8) invariant reformulation of $D=11$ supergravity \cite{dWNsu8}, 
has been constructed. The central object in this reformulation is an E$_{7(7)}$ 56-bein
{\em in eleven dimensions}, which can be thought of as the eleven dimensional ancestor 
of the 56-bein in four dimensions containing the 70 scalars of the reduced 
maximal theory.  The four generalised vielbeine \cite{dWNsu8, dWN13, GGN13} that  
comprise the 56-bein in eleven dimensions are derived by 
analysing the supersymmetry transformations of the 56 vector fields in the SU(8) invariant reformulation, generalising and completing the construction of \cite{dWNsu8}
(similar new structures also appear in the SO(16) invariant formulation of $D=11$ supergravity 
where the relevant vielbein belongs to E$_{8(8)}$ \cite{Nso16,KNS}). 
The emphasis on supersymmetry as the origin of the generalised exceptional geometry
obtained in this way is the main distinctive feature in comparison with other approaches
to generalised geometry.~\footnote{For a summary of recent developments 
and a complete bibliography see \cite{btreview, Hohm:2013vpa}.} The 56-bein satisfies
certain differential identities called `generalised vielbein postulates' \cite{dWNsu8, GGN13} 
due to their similarities with the usual vielbein postulate in differential geometry, and these
relations will be at the center of our construction.

The very nature of the reformulation in that it emphasises structures in eleven dimensions that become apparent upon reduction to four dimensions makes it a useful framework in which to study questions regarding four-dimensional maximal gauged theories from a higher dimensional perspective.  This feature extends the attributes of the SU(8) invariant reformulation, which leads to a non-linear metric ansatz \cite{dWNW} and a proof \cite{dWNconsis, NP} of the consistency of the $S^7$ reduction \cite{duffpope} of $D=11$ supergravity.  In particular, the new structures found in \cite{dWN13, GGN13} give rise to non-linear ans\"atze for the internal components of the three-form \cite{dWN13} (see also \cite{GGN}) and six-form \cite{KKdual} potentials.  In fact, ans\"atze can be given for the full uplift to eleven dimensions for any solution (and, in particular, the
stationary points of the potential) of the four-dimensional theory; 
the possibility to perform such non-trivial tests of all
formulae is another distinctive feature of the present approach.
Furthermore, the generalised vielbein postulates reduce to the consistency requirements of the four dimensional maximal gauged theory.  In particular,  there is a direct relation \cite{GGN13, KKdual} between the set of generalised vielbein postulates with derivatives along four dimensions and the E$_{7(7)}$ Cartan equation of the maximal gauged theory \cite{dWSTlag, dWSTgauge, dWSTmax4}, in which the gauging is defined via the embedding tensor \cite{NSmaximal3, NScomgauge3, dWSTlag}. 

The formalism developed in \cite{GGN13} has already been applied to an extensive study of the $S^7$ reduction \cite{KKdual}.  In particular, nonlinear ans\"atze are given for the uplift of four-dimensional solutions of SO(8) gauged maximal supergravity \cite{dWNn8} to eleven dimensions, including dual fields.  In addition, the embedding tensor of SO(8) gauged maximal supergravity is recovered directly by reducing the generalised vielbein postulates with derivatives along four dimensions.  While the $S^7$ reduction is highly non-trivial from the perspective of the non-linearity of uplift ans\"atze and the field content in four dimensions, the gauging, and therefore the embedding tensor, is relatively simple in that the gauging only involves electric vectors, and moreover is uniform.

In this paper, we study Scherk-Schwarz \cite{scherkschwarz}~\footnote{In fact, the essential idea of reducing on a group manifold appears in \cite{dewitt};  for a useful historical account of 
Kaluza-Klein theory see \cite{CGP}.} 
reductions of $D=11$ supergravity with background flux \cite{ttf1, ttf2, DAF, Mtwistedtorus, ttf3, dallprez, Fre, ssdft, fretrig, ttf4, SamLect} within the context of the formalism developed in \cite{GGN13}.  The Scherk-Schwarz flux compactification has principally been studied from a four-dimensional gauge algebra perspective by associating background fields to particular representations in the GL(7) decomposition of the {\bf 912} representation of E$_{7(7)}$ in 
which the embedding tensor lives.  Here, we concentrate on obtaining the embedding tensor 
of such theories {\em directly from eleven dimensions} by analysing the couplings of the 56 vector
fields (28 electric and 28 magnetic vectors) via the generalized vielbein postulates. Hence,
our approach should be contrasted with recent work \cite{Grana:2012rr, Berman:2012uy, Musaev:2013rq, Aldazabal:2013mya} aiming to construct the embedding tensor for non-geometric compactifications obtained by generalised Scherk-Schwarz reductions of extended 
generalised geometries.

While the Scherk-Schwarz reduction is much simpler than the $S^7$ reduction, the novelty of the Scherk-Schwarz reduction as far as we are interested in is the potential for gaugings involving a combination of electric and magnetic vectors leading to a more complicated embedding tensor \cite{Mtwistedtorus, ssdft}.  We derive the embedding tensor of Scherk-Schwarz flux compactifications directly and explicitly from the $D=11$ generalised vielbein postulates. This constitutes a further non-trivial demonstration of the utility of the formalism developed in Ref.~\cite{GGN13} and gives further credence to the interpretation of the generalised vielbein postulates as the higher dimensional origin of the embedding tensor.  More generally, the results of Ref.~\cite{GGN13} can be applied to any compactification of $D=11$ supergravity to maximal gauged theories in four dimensions yielding non-linear uplift ans\"atze and the embedding tensor.

The outline of the paper is as follows: In section \ref{sec:ss}, we present a self-contained review of Scherk-Schwarz reductions with background flux including a discussion of the background field equations (section \ref{sec:ssbgd}), which to the best of our knowledge does not appear in previous literature.  The Jacobi-like constraints on the background fluxes as well as the background field equations form the complete set of equations that must be satisfied for a {\textit{bona fide}} Scherk-Schwarz flux compactification. The non-triviality of these constraints, particularly the background field equations, illustrates the difficulty of providing a complete classification of such compactifications.  

In section \ref{sec:embedding}, we briefly review the embedding tensor formalism \cite{NSmaximal3, NScomgauge3, dWSTlag, dWSTgauge, dWSTmax4} and give a general solution of the linear constraint satisfied by the embedding tensor.  The reduction ans\"atze defined in section \ref{sec:ss} are applied to the generalised vielbein postulates in section \ref{sec:gvp} yielding the embedding tensor of Scherk-Schwarz flux compactifications.  This embedding tensor can be cast in the form of the general solution of the linear constraint given in section \ref{sec:embedding}.  Furthermore, in appendix \ref{app:quad}, we verify that the quadratic constraints are satisfied.  Finally, in section \ref{sec:flat}, we demonstrate explicitly in the simple example of a flat group reduction that indeed less than or equal to 28 electric or magnetic vectors are gauged as is expected from general results of the embedding tensor formalism \cite{SamLect}.  We make concluding remarks in section \ref{sec:con}.

{\bf Conventions:} In this paper, we reserve the use of $\epsilon$ for an alternating \emph{tensor} with respect to some metric structure, while we use $\eta$ to denote the {\em tensor density},
alias the alternating symbol.  It is important to note that \emph{all} objects denoted 
with a caret ($\hat{\ \, }$) above them depend only on the external coordinates, that is,
are only $x$-dependent.

\section{Scherk-Schwarz reduction} \label{sec:ss}

Consider a reduction of $D=11$ supergravity such that the elfbein takes the form
\begin{equation} \label{elfbein}
 {E_M}^A (z) = \begin{pmatrix}
                 \hat{\Delta}^{-1/2}(x)\, {\hat{e}_{\mu}}{}^{\alpha}(x) & {\hat{B}_{\mu}}{}^m(x)\, {\hat{e}_m}{}^a(x) \\[3mm]
		    0 & {U_m}^{n}(y)\, {\hat{e}_n}{}^a(x)
                \end{pmatrix},
\end{equation}
where the eleven dimensional coordinates have been split as
$\{z^M\} \equiv \{ x^\mu, y^m\}$, and where
\begin{equation}
 \hat{e}= \det ({\hat{e}_{\mu}}{}^{\alpha}), \qquad \hat{\Delta}= \det ({\hat{e}_m}{}^a)
\end{equation}
(recall that all hatted quantities depend only on the  four-dimensional coordinates $x^\mu$).
The matrices ${U_m}^{n}(y)$ depend only on the internal coordinates and satisfy the property that
\begin{equation} \label{dU}
 \partial_{[m} {U_{n]}}^p = - \frac{1}{2} {f^{p}}_{rs} {U_m}^{r} {U_n}^{s}.
\end{equation}
The $y$-independent structure constants $f$ importantly satisfy a unimodularity property, \textit{viz.}
\begin{equation} \label{funi}
 {f^m}_{mn}=0,
\end{equation}
which is equivalent to
\begin{equation} \label{Uunimod}
 \partial_n \left[ U {(U^{-1})_{m}}^n \right]= 0,
\end{equation}
where
\begin{equation}
 U \equiv \text{det}({U_m}^{n}).
\end{equation}
The condition of unimodularity, emphasised in \cite{scherkschwarz} ensures that the measure is invariant under seven-dimensional diffeomorphisms.~\footnote{The importance of unimodularity 
was discussed in the context of Bianchi cosmology by Sneddon \cite{sneddon} 
slightly before Scherk and Schwarz, and shown to be required for consistency of the
reduction to a homogeneous cosmology.}   

Furthermore, the following integrability condition is satisfied
\begin{equation} \label{ffjac}
 f^{q}{}_{[mn} f^{r}{}_{p]q}=0.
\end{equation}
This is equivalent to the Jacobi identity for the associated Lie algebra.

Specifically, in terms of the following parametrisation of the elfbein
\begin{equation} \label{elfbeinorig}
 {E_M}^A = \begin{pmatrix}
                 \Delta^{-1/2} e'_{\mu}{}^{\alpha} & B_{\mu}{}^n e_{n}{}^a \\[3mm]
		    0 & e_{m}{}^a
                \end{pmatrix},
\end{equation}
where $\Delta= \det e_{n}{}^{a} = U \hat{\Delta},$
 we assume the following reduction ans\"atze for the elfbein components
\begin{align}
 {e'_\mu}{}^\alpha(x,y) &= U^{1/2} {\hat{e}_{\mu}}{}^{\alpha}(x), \\
 {B_{\mu}}^m(x,y) &= {{(U^{-1})}_n}^m {\hat{B}_{\mu}}{}^n(x), \label{ansBvec} \\
 {e_n}^a(x,y) &= {U_m}^{n} {\hat{e}_n}{}^a(x). \label{anssieb}
\end{align}
In general, the reduction ans\"atze for fields is such that all seven-dimensional covariant tensor indices are contracted with $U$, which contains all the $y$-dependence, while seven-dimensional contravariant tensor indices are contracted with $U^{-1}$, as should be clear from the ans\"atze for ${B_{\mu}}^m$ and ${e_n}^a$ given above.

The reduction ansatz for the 3-form potential is similarly defined, except that some components have background contributions as well.
\begin{align}
A_{\mu \nu \rho}(x,y) &= \hat{A}_{\mu \nu \rho}(x) + \hat{\zeta}_{\mu \nu \rho}(x), \label{Aans:1} \\
A_{\mu \nu m}(x,y) &= {U_{m}}^n \hat{A}_{\mu \nu n}(x), \\
A_{\mu m n}(x,y) &=  {U_{m}}^p {U_{n}}^q \hat{A}_{\mu p q}(x), \\
A_{m n p}(x,y) &= A'_{m n p}(x,y) + a_{m n p}(y),
\end{align}
where
\begin{equation}
 A'_{m n p}(x,y) = {U_{m}}^q {U_{n}}^r {U_{p}}^s \hat{A}_{q r s}(x)
\end{equation}
and $\hat{\zeta}_{\mu \nu \rho}$ and $a_{m n p}$ are defined such that
\begin{align}
4! \partial_{[\mu} \hat{\zeta}_{\nu \rho \sigma]} &=i \fr \hat{\Delta}^{-3} \hat{\epsilon}_{\mu \nu \rho \sigma}, \label{zeta3} \\
4! \partial_{[m} a_{npq]} &= g_{rstu} {U_{m}}^r {U_{n}}^s {U_{p}}^t {U_{q}}^u, \label{da3}
\end{align}
for some constant $\fr$ and totally antisymmetric constant $g_{mnpq}$.  The above equations 
give the background values of the field strength $F_{\mu \nu \rho \sigma}$ and $F_{mnpq}$, respectively. We will see later that the special $y$-dependence with {\em constant} $g_{mnpq}$ in
\eqref{da3} is required for the consistency of both the equations of motion and the
generalised vielbein postulates.

The exterior derivative of equation \eqref{da3}, which corresponds to the closure of the background field strength, implies the following constraint \cite{ttf4}
\begin{equation} \label{fgjac}
{f^s}_{[mn}g_{pqr]s}=0.
\end{equation}
We will find later that this constraint plays a crucial role in defining a consistent gauge algebra.  In fact, this constraint was first found by considering the consistency of the gauge algebra, in particular the Jacobi identity \cite{ttf1}.

In order to determine the form of the dual six-form under this reduction, we consider its defining equation
\begin{equation} \label{dualityrel}
  \frac{i}{4!} \epsilon_{M_1 \ldots M_{11}} F^{M_8 \ldots M_{11}}= 7! \partial_{[M_1} A_{M_2 \dots M_7]} + 7! \frac{\sqrt{2}}{2}  A_{[M_1 \dots M_3} \partial_{M_4} A_{M_5 \dots M_7]},
\end{equation}
where it is important to note that indices on $F^{MNPQ}$ have been raised using the eleven-dimensional metric, and where we have ignored fermion bilinear contributions.
Consider the $m_1 \ldots m_7$ components of the above equation.  Using the fact that 
\begin{equation}
 \epsilon_{m_1 \ldots m_7 \mu \nu \rho \sigma}=U \hat{\Delta}^{-1} \hat{\epsilon}_{\mu \nu \rho \sigma} \eta_{m_1 \ldots m_7}
\end{equation}
the left hand side of equation \eqref{dualityrel} simplifies to
\begin{equation} \label{dualityrelcom}
 \frac{i}{4!} \epsilon_{m_1 \ldots m_{7} \mu \nu \rho \sigma} F^{\mu \nu \rho \sigma}= - U \fr \eta_{m_1 \ldots m_7} + U(\text{$x$-dependent terms}),
\end{equation}
where $\eta_{m_1 \ldots m_7}$ is defined with respect to a flat seven-dimensional metric and the $x$-dependent terms in the remainder of the expression have contributions from $\hat{A}_{\mu \nu \rho}$, $\hat{A}_{\mu \nu m}$, $\hat{A}_{\mu mn}$, $\hat{A}_{mnp}$ and $g_{mnpq}$ as well as ${f^p}_{mn}$.  This is due to the fact that the inverse metric is not diagonal. We stress once more that the indices on the 4-form $F$ in equation \eqref{dualityrelcom} have been raised with the eleven-dimensional metric.

The right hand side of equation \eqref{dualityrel} reduces to
\begin{align}
7! \partial_{[m_1}& \left( A_{m_2 \ldots m_7]} + \frac{\sqrt{2}}{2} A'_{m_2 m_3 m_4} a_{m_5 m_6 m_7]} \right) \notag \\[2mm]
 &\hspace{10mm}+\frac{7!\sqrt{2}}{2} \left( A'_{[m_1 m_2 m_3} \partial_{m_4} (A'_{m_5 m_6 m_7]} + 2 a_{m_5 m_6 m_7]})
+ a_{[m_1 m_2 m_3} \partial_{m_4} a_{m_5 m_6 m_7]} \right).
\label{def:dualform}
\end{align}

Now, defining an ansatz for $A_{m_1 \ldots m_6}$ of the form
\begin{equation}
 A_{m_1 \ldots m_6} = A'_{m_1 \ldots m_6}(x,y) + \frac{\sqrt{2}}{2} a_{[m_1 m_2 m_3} A'_{m_4 m_5 m_6]} + a_{m_1 \ldots m_6}(y),
\end{equation}
where
\begin{equation}
 A'_{m_1 \ldots m_6}(x,y) = {U_{m_1}}^{n_1} \dots {U_{m_6}}^{n_6} \hat{A}_{n_1 \ldots n_6}(x)
\end{equation}
and $a_{m_1 \ldots m_6}$ is such that
\begin{equation} \label{bgddual}
7! \partial_{[m_1} a_{m_2 \ldots m_7]} =  - U \fr \eta_{m_1 \ldots m_7} - \frac{7!\sqrt{2}}{2} a_{[m_1 m_2 m_3} \partial_{m_4} a_{m_5 m_6 m_7]},
\end{equation}
equation \eqref{dualityrel} reduces to a purely $x$-dependent, rather complicated, relation between $\hat{A}_{m_1 \ldots m_6}$ and components of the three-form potential $\hat{A}$.  Note that duality relation \eqref{bgddual} is the duality relation satisfied by the background solution.

\subsection{Background solution} \label{sec:ssbgd}

In the context of formulating a well-defined reduction, an important consideration is the background field equations and the constraints these imply on the background fields.

The background of the Scherk-Schwarz reduction is given by 
\begin{equation}
  \label{backgroundelfbein}
 {E_M}^A = \begin{pmatrix}
                 \hat{e}_{\mu}{}^{\alpha}(x) & 0 \\
		    0 & {U_m}^{n}(y)\, \delta_{n}^a(x)
                \end{pmatrix}, \qquad A_{mnp}= a_{mnp}, \qquad F_{\mu \nu \rho \sigma} = i \fr \hat{\epsilon}_{\mu \nu \rho \sigma}.
\end{equation}
Thus, the internal metric is
\begin{equation}
 g_{mn}= {U_m}^{p} {U_n}^{q} \delta_{pq}, \qquad g^{mn} = (U^{-1})_{p}{}^{m} (U^{-1})_{q}{}^{n} \delta^{pq}.
\end{equation}

The field equations of eleven-dimensional supergravity are
\begin{gather}
 R_{MN} = \textstyle{\frac{1}{72}} g_{MN} F^2_{PQRS} - \textstyle{\frac{1}{6}} F_{MPQR} {F_N}^{PQR}, \label{11dEinstein} \\[7pt]
 E^{-1} \partial_{M} (E F^{MNPQ}) = \textstyle{\frac{\sqrt{2}}{1152}} i \epsilon^{NPQR_1\ldots R_4 S_1 \ldots S_4} F_{R_1\ldots R_4} F_{S_1\ldots S_4}. \label{11dMaxwell}
\end{gather}
For the background solution, the component of these equations along the internal directions are
\begin{gather}
 \frac{1}{6} g_{mpqr} g_n{}^{pqr} = \frac{1}{4} \left(\delta_{m p} \delta_{n q} \delta^{rs} \delta^{tu} f^{p}{}_{r t} f^{q}{}_{s u} - 2 \delta_{p q} \delta^{r s} f^{p}{}_{m r} f^{q}{}_{n s} - 2 f^{p}{}_{m q} f^{q}{}_{n p} \right)  \notag \\[4pt] \hspace{100mm} - \frac{1}{3} \delta_{mn} \fr^{2} + \frac{1}{72} \delta_{mn} g_{pqrs} g^{pqrs}, \label{bkg:einstein} \\[4pt]
{f^{[m_1}}_{pq} g^{m_2 m_3]pq} = - \frac{\sqrt{2}}{72} \fr \eta^{m_1 \ldots m_7} g_{m_4 \ldots m_7}, \label{bkg:maxwell}
\end{gather}
where the indices on $g_{mnpq}$ are raised with the Kronecker $\delta$ symbol. We note that
by putting the theory on-shell, this operation breaks the GL(7) symmetry to SO(7) or a subgroup thereof, in the same way as the rigid SU(8) symmetry of maximal supergravity is broken 
to (a subgroup of) SO(8) in any given vacuum.\footnote{We thank 
Henning Samtleben for a discussion on this point.}  The special
dependence on $U(y)$ in \eqref{da3} is now seen to be necessary for the `Maxwell equation'
\eqref{11dMaxwell} to become $y$-independent, and thus
to reduce to an equation relating the {\em constant} tensors $f^m{}_{np}$
and $g_{mnpq}$, \eqref{bkg:maxwell}. We note that, while the background constraints for the case with no flux 
appear already in Ref.~\cite{scherkschwarz}, the constraints implied on the background 
of a Scherk-Schwarz reduction with flux have never been fully spelled out in the literature 
to the best of our knowledge.  In particular, equation \eqref{bkg:maxwell} is a non-trivial restriction on the class of viable Scherk-Schwarz reductions.  These constraints, which are imposed by the background field equations are independent of the constraints imposed by the consistency of the gauge algebra \cite{ttf1} (see also \cite{SamLect}).  

The components of the Einstein equation along the four-dimensional spacetime directions fixes the radius of the four-dimensional Anti-de Sitter space
\begin{equation} \label{bkg:4d}
 {\hat{R}_{\mu}}{}^{\nu} = \left(\textstyle{\frac{2}{3}} \fr^2 + \textstyle{\frac{1}{72}} g^{mnpq}g_{mnpq} \right) \delta^{\nu}_{\mu}.
\end{equation}
All other equations of motion are trivially satisfied. 

\section{The embedding tensor formalism} \label{sec:embedding}

The embedding tensor formalism~\footnote{See Ref.~\cite{SamLect} for a lucid account of the embedding tensor formalism.}, which was initially developed in the context of three-dimensional maximal gauged supergravities \cite{NSmaximal3, NScomgauge3} and later developed in the context of four-dimensional maximal gauged supergravities \cite{dWSTlag, dWSTgauge, dWSTmax4} is the most efficient framework in which to understand gaugings.  The embedding tensor formalism uses the fact that the ungauged supergravity, of which the gauged theory is a deformation, is controlled by a global symmetry group that is larger than what one would naively expect---an observation first made in the context of the four dimensional maximal theory \cite{cremmerjulia}.

In four dimensions, the scalars, which parametrise the E$_{7(7)}$ vielbein $\cV$ satisfy the following equation
\begin{equation} \label{embedeqn}
 \partial_{\mu} \cV_{\cM ij} + \cQ_{\mu}{}^{k}{}_{[i} \cV_{\cM \, j]k} - \cP_{\mu ijkl} \cV_{\cM}{}^{kl} - g \cA_{\mu}{}^{\cP} X_{\cP \cM}{}^{\cN} \cV_{\cN ij}=0.
\end{equation}

Objects that are of particular interest in the above equation are $(X_{\cM})_{\cN}{}^{\cP}$. These generate the gauge algebra and are related to the embedding tensor~\footnote{Indices $\alpha, \beta, \ldots=1,\ldots,133$ label the E$_{7(7)}$ generators and are not to be confused with the four-dimensional tangent space indices, which are also labelled by lower Greek letters from the beginning of the alphabet.} $\Theta_{\cM}{}^{\alpha}$ via the E$_{7(7)}$ generators $t_{\alpha}$, {\it{viz.}}
\begin{equation}
X_{\cM} = \Theta_{\cM}{}^\alpha\,  t_{\alpha}.
\end{equation}

The embedding tensor satisfies two algebraic constraints.  The first, linear constraint, comes from a consideration of the supersymmetric consistency of the gauged theory.  In the case of maximal four-dimensional theories, this translates to the statement that the embedding tensor lives in the {\bf 912} representation of E$_{7(7)}$
\begin{equation}
 \Theta_{\cM}{}^{\alpha} + 2 (t_\beta)_{\cM}{}^{\cN} (t^\alpha)_{\cN}{}^{\cP} \Theta_{\cP}{}^{\beta}=0,
\end{equation}
where the E$_{7(7)}$ index $\alpha$ is raised with the inverse Killing-Cartan form $\kappa^{-1}$, which is given in appendix \ref{app:e7}. More specifically, the above relation follows by
requiring that the projectors  ${\mathbb{P}}_{\bf{56}}$ and ${\mathbb{P}}_{\bf{6480}}$
annihilate $\Theta_{\cM}{}^\alpha$ \cite{dWSTlag}.
In terms of the gauge group generators, the linear constraint is
\begin{equation}
 X_{\cM \, \cN}{}^{\cP} + 2X_{\cR \, \cM}{}^{\cQ} (\kappa^{-1})^{\alpha \beta} (t_\alpha)_{\cQ}{}^{\cR} (t_{\beta})_{\cN}{}^{\cP} = 0.
\end{equation}
The general solution of the linear constraint is given by
\begin{align}
 X_{\tM \tN}{}^{\tP\tQ}{}_{\tR\tS} &= \delta^{[\tP}_{[\tR}\tT^{\tQ]}{}_{\tS]\tM\tN}, 
 \hspace{27.8mm} X_{\tM\tN \, \tP\tQ}{}^{\tR\tS} = - \delta^{[\tR}_{[\tP}\tT^{\tS]}{}_{\tQ]\tM\tN}, \notag \\[2mm]
X^{\tM\tN}{}_{\tP\tQ\, \tR\tS} &= -2 \delta^{[\tM}_{[\tP}\tT^{\tN]}{}_{\tQ\tR\tS]},
\hspace{22.15mm} X^{\tM\tN\, \tP\tQ\, \tR\tS} = - \frac{2}{4!} \eta^{\tP\tQ\tR\tS[\tM|\tT_1 \tT_2 \tT_3|} \tT^{\tN]}{}_{\tT_1 \tT_2 \tT_3}, \notag \\[2mm]
 X^{\tM\tN}{}_{\tP\tQ}{}^{\tR\tS} &= \delta^{[\tR}_{[\tP} \tT_{\tQ]}{}^{\tS]\tM\tN}, 
\hspace{28mm} X^{\tM\tN\, \tP\tQ}{}_{\tR\tS} = - \delta^{[\tP}_{[\tR} \tT_{\tS]}{}^{\tQ]\tM\tN}, \notag \\[2mm]
 X_{\tM\tN}{}^{\tP\tQ \, \tR\tS} &= -2 \delta^{[\tP}_{[\tM} \tT_{\tN]}{}^{\tQ\tR\tS]},
\hspace{23mm} X_{\tM\tN \, \tP\tQ \, \tR\tS} = - \frac{2}{4!} \eta_{\tP\tQ\tR\tS[\tM|\tT_1 \tT_2 \tT_3|} \tT_{\tN]}{}^{\tT_1 \tT_2 \tT_3},
  \label{X}
\end{align}
where
\begin{equation}
  \tT^{\tM}{}_{\tN\tP\tQ} = -\frac{3}{4} {\tA_2}^{\tM}{}_{\tN\tP\tQ} - \frac{3}{2} \delta^{\tM}_{[\tP}\tA_{1\, \tQ] \tN}, \qquad
 \tT_{\tM}{}^{\tN\tP\tQ} = -\frac{3}{4} \tA_{2\, \tM}{}^{\tN\tP\tQ} - \frac{3}{2} \delta_{\tM}^{[\tP}{\tA_{1}}^{\tQ] \tN}.
\end{equation}
Note that the solution above applies more generally to other compactifications.  Structures $\tA_{1\, \tM \tN}$, ${\tA_{1}}^{\tM \tN}$ ${\tA_2}^{\tM}{}_{\tN\tP\tQ}$ and $\tA_{2\, \tM}{}^{\tN\tP\tQ}$ satisfy the following properties
\begin{gather}
 \tA_{1\, [\tM\tN]} =0, \qquad {\tA_{1}}^{[\tM \tN]}=0, \notag \\
 {\tA_2}^{\tM}{}_{[\tN\tP\tQ]} = {\tA_2}^{\tM}{}_{\tN\tP\tQ}, \qquad   {\tA_2}^{\tM}{}_{\tM\tP\tQ}=0, \notag \\
 \tA_{2 \, \tM}{}^{[\tN\tP\tQ]} = \tA_{2 \, \tM}{}^{\tN\tP\tQ}, \qquad \tA_{2 \, \tM}{}^{\tM\tP\tQ}=0.
\end{gather}

Equivalently,
\begin{align}
 (\Theta_{\tM \tN})_{\tP_1}{}^{\tP_2}&=\frac{1}{2} \tT^{\tP_2}{}_{\tP_1 \tM \tN}, 
\hspace{18.5mm} (\Theta^{\tM \tN})^{\tP_1 \ldots \tP_4} = -\frac{2}{4!} \eta^{\tP_1 \ldots \tP_4 [\tM|\tQ_1 \tQ_2 \tQ_3|} \tT^{\tN]}{}_{\tQ_1 \tQ_2 \tQ_3}, \notag \\
 (\Theta^{\tM \tN})_{\tP_1}{}^{\tP_2}&=-\frac{1}{2} \tT_{\tP_1}{}^{\tP_2 \tM \tN},
\qquad \qquad (\Theta_{\tM \tN})^{\tP_1 \ldots \tP_4} = -2 \delta_{[\tM}^{[\tP_1} \tT_{\tN]}{}^{\tP_2 \tP_3 \tP_4]}.
\end{align}
The corresponding objects $(\Theta^{\tM \tN})_{\tP_1 \ldots \tP_4}$ and $(\Theta_{\tM \tN})_{\tP_1 \ldots \tP_4}$ are obtained by contracting $(\Theta^{\tM \tN})^{\tP_1 \ldots \tP_4}$ and $(\Theta_{\tM \tN})^{\tP_1 \ldots \tP_4}$ with the permutation symbol in accordance with the equations in appendix \ref{app:e7}.

It is important to note at this point that $\tT^{\tM}{}_{\tN\tP\tQ}$ and $\tT_{\tM}{}^{\tN\tP\tQ}$ are real 
and completely independent.  This is because they are written in terms of SL(8) indices and there is no relation between an upper SL(8) index and a lower one.  This is in contrast to objects with SU(8) indices where upper and lower indices are related to one another via conjugation.  The $T$-tensor, which has SU(8) indices can be derived by dressing the $\tT$-tensors above with the E$_{7(7)}$ vielbein $\cV_{\cM\, ij}$
\begin{equation}
 T_{i_1 i_2}{}^{j_1 j_2}{}_{k_1 k_2} = - \Omega^{\cM \cQ} \Omega^{\cN \cR} \cV_{\cQ\, i_1 i_2} \cV_{\cR}{}^{j_1 j_2} \cV_{\cP\, k_1 k_2} X_{\cM\, \cN}{}^{\cP},
\end{equation}
where
\begin{equation}
 T_{i_1 i_2}{}^{j_1 j_2}{}_{k_1 k_2} = \delta^{[j_1}_{[i_1} T^{j_2]}{}_{i_2] k_1 k_2}
\end{equation}
and
\begin{equation}
 T^{i}{}_{jkl} = -\frac{3}{4} {A_2}^{i}{}_{jkl} - \frac{3}{2} \delta^{i}_{[k}A_{1\, l] j}.
\end{equation}
Since the $T$-tensor has SU(8) indices, $T_{i}{}^{jkl}$ is simply the complex conjugate 
of $T^{i}{}_{jkl}$.  Note that this is in contrast to the properties satisfied by the $\tT$-tensors which satisfy no such relation, as pointed out above.

Furthermore, the embedding tensor satisfies a quadratic constraint, which is necessary for the gauge algebra generated by $X_{\cM}$ to close
\begin{equation} \label{cons:quad}
 X_{\cM \cQ}{}^{\cR} X_{\cN \cR}{}^{\cP} - X_{\cN \cQ}{}^{\cR} X_{\cM \cR}{}^{\cP} + X_{\cM \cN}{}^{\cR} X_{\cR \cQ}{}^{\cP}=0.
\end{equation}
However, notice that the above constraint is stronger than the closure of the algebra since $X_{(\cM \cN)}{}^{\cP}$ does not trivially vanish.  In fact, the quadratic condition comes from the requirement that the embedding tensor be invariant under the action of the gauge group
\begin{equation}
 \delta_{\cM} \Theta_{\cN}{}^{\alpha} = \Theta_{\cM}{}^{\beta} \delta_{\beta} \Theta_{\cN}{}^{\alpha} = 0.
\end{equation}

Equivalently, given that the embedding tensor satisfies the linear constraint and lives in the {\bf 912} representation of E$_{7(7)}$, the quadratic constraint is \cite{SamLect}
\begin{equation}
 \Omega^{\cM \cN} \Theta_{\cM}{}^{\alpha} \Theta_{\cN}{}^{\beta}=0. 
\end{equation}
In this form, it is clear to see that viewed as a matrix, the row rank of the embedding tensor is at most half-maximal.  Therefore, we are guaranteed that only at most 28 out of the possible 56 vectors will be gauged \cite{SamLect}.

\section{Generalised vielbein postulates and the embedding tensor} \label{sec:gvp}

The generalised vielbein postulates provide an understanding of various aspects of the reduction.  In particular, for the case of the $S^7$ compactification, they are a necessary ingredient in the proof of the consistency of the reduction.  Specifically, the $d=4$ generalised vielbein postulates reduce to the E$_{7(7)}$ Cartan equation of gauged maximal supergravity in that case \cite{dWNconsis, KKdual}.

The generalised vielbeine combine the would-be scalar degrees of freedom originating
from the siebenbein, the 3-form and the 6-form into a single object, and are explicitly
given by \cite{GGN13}:
\begin{align}
e^{m}_{AB} &= i\Delta^{-1/2} \Gamma^m_{AB}, \label{gv1} \\[3mm]
e_{mn}{}_{AB} &= - \frac{\sqrt{2}}{12} i \Delta^{-1/2} \left(
\Gamma_{mn}{}_{AB} + 6 \sqrt{2} A_{mnp} \Gamma^{p}_{AB} \right),
\label{gv2} \\[3mm]
e_{m_1 \dots m_5}{}_{AB}&= \frac{1}{6!\sqrt{2}} i \Delta^{-1/2} \Bigg[
\Gamma_{m_1 \dots m_5}{}_{AB} + 60 \sqrt{2} A_{[m_1 m_2 m_3} \Gamma_{m_4
m_5]}{}_{AB} \notag \\
 & \hspace*{50mm}  - 6! \sqrt{2} \Big( A_{p m_1 \dots m_5} - 
\frac{\sqrt{2}}{4} A_{p[m_1 m_2} A_{m_3 m_4 m_5]} \Big) \Gamma^{p}_{AB}
\Bigg], \label{gv3} \\[2mm]
e_{m_1 \dots m_7, n}{}_{AB} &=- \frac{2}{9!} i \Delta^{-1/2} \Bigg[
(\Gamma_{m_1 \dots m_7} \Gamma_{n}{})_{AB} + 126 \sqrt{2}\ A_{n [m_1
m_2} \Gamma_{m_3 \dots m_7]}{}_{AB} \notag \\
 & \hspace*{37mm}  + 3\sqrt{2} \times 7! \Big( A_{n [ m_1 \dots m_5} +
\frac{\sqrt{2}}{4} A_{n[m_1 m_2} A_{m_3 m_4 m_5} \Big) \Gamma_{m_6
m_7]}{}_{AB} \notag \\[2mm]
  & \hspace*{42.5mm} + \frac{9!}{2} \Big(A_{n [ m_1 \dots m_5} +
\frac{\sqrt{2}}{12} A_{n[m_1 m_2} A_{m_3 m_4 m_5} \Big) A_{m_6 m_7] p }
\Gamma^{p}{}_{AB} \Bigg]. \label{gv4}
\end{align}
We emphasize again that these objects depend on all eleven coordinates.
By virtue of their definition, they satisfy certain differential constraints, the so-called
generalised vielbein postulates. Along the external $d=4$ directions these
are of the form
\begin{align} \label{cDB1}
 & \cD_{\mu} e^{m}_{AB} + \frac{1}{2} \partial_{n} B_{\mu}{}^{n} e^{m}_{AB} + \partial_{n} B_{\mu}{}^{m} e^{n}_{AB} + \cQ_{\mu}^{C}{}_{[A} e^{m}_{B]C} + \cP_{\mu ABCD} e^{mCD} = 0, \\[3mm]
\label{cDB2}
&\cD_\mu e_{mnAB} + \frac{1}{2} \partial_{p} B_{\mu}{}^{p} e_{mnAB} + 2 \partial_{[m} 
B_{|\mu|}{}^{p} e_{n]pAB} + 3 \partial_{[m} B_{|\mu|np]} e^{p}_{AB} \notag \\[2pt]
&\hspace{80mm}+ \cQ_{\mu}^{C}{}_{[A} e_{mnB]C} + \cP_{\mu ABCD} e_{mn}{}^{CD} = 0, \\[3mm]
&\cD_\mu e_{m_1 \dots m_5 AB} + \frac{1}{2} \partial_{p} B_{\mu}{}^{p} e_{m_1 \dots m_5 AB} - 5 \partial_{[m_1} B_{|\mu|}{}^{p} e_{m_2 \dots m_5]pAB} + 
\frac{3}{\sqrt{2}} \partial_{[m_1} B_{|\mu|m_2 m_3} e_{m_4 m_5]AB} \notag \\[3pt] \label{cDB3}
& \hspace{40mm}- 6 \partial_{[m_1} B_{|\mu| m_2 \dots m_5 p]} e^{p}_{AB} + \cQ_{\mu}^{C}{}_{[A} e_{m_1 \dots m_5 B]C} + \cP_{\mu ABCD} e_{m_1 \dots m_5}{}^{CD}=0, \\[10pt]
& \cD_\mu e_{m_1 \dots m_7, n AB} - \frac{1}{2} \partial_{p} B_{\mu}{}^{p} e_{m_1 \dots m_7, n AB} - \partial_{n} B_{\mu}{}^{p} e_{m_1 \dots m_7, p AB} + 5 \partial_{[m_1} B_{|\mu| m_2 m_3} e_{m_4 \dots m_7] n AB}  \notag \\[2mm] 
& \hspace{30mm} - 2 \partial_{[m_1} B_{|\mu| m_2 \dots m_6} e_{m_7] n AB}+ \cQ_{\mu}^{C}{}_{[A} e_{m_1 \dots m_7, n B]C} + \cP_{\mu ABCD} e_{m_1 \dots m_7, n}{}^{CD} =0, \label{cDB4}
\end{align}
where
\begin{equation}
\cD_\mu \equiv \partial_\mu - B_\mu{}^m \partial_m.
\end{equation}
and the connection coefficients are of the form
\begin{align}
 \cQ_{\mu}^{A}{}_{B} &= - \textstyle{\frac{1}{2}} \left[ {e^m}_a \partial_{m} B_{\mu}{}^{n} e_{n b} - ({e^p}_{a} \cD_{\mu} e_{p\, b}) \right] \Gamma^{ab}_{AB} 
 - \textstyle{\frac{\sqrt{2}}{12}} {e_{\mu}}{}^{\alpha} \left( F_{\alpha abc} \Gamma^{abc}_{AB} - \eta_{\alpha \beta \gamma \delta} F^{\beta \gamma \delta a} \Gamma_{a AB} \right), \\[3mm]
 \cP_{\mu ABCD}& = \textstyle{\frac{3}{4}} \left[ {e^m}_a \partial_{m} B_{\mu}{}^{n} e_{n b} - ({e^p}_{a} \cD_{\mu} e_{p\, b}) \right] \Gamma^{a}_{[AB} \Gamma^{b}_{CD]}
 - \textstyle{\frac{\sqrt{2}}{8}} {e_{\mu}}{}^{\alpha} F_{abc \alpha} \Gamma^{a}_{[AB} \Gamma^{bc}_{CD]} \notag \\[2mm]
& \hspace{75mm} - \textstyle{\frac{\sqrt{2}}{48}} e_{\mu \, \alpha} \eta^{\alpha \beta \gamma \delta} F_{a \beta \gamma \delta}{\Gamma_{b}}_{[AB} \Gamma^{ab}_{CD]}.
\end{align}
Below we will consider and analyse these equations in the context of Scherk-Schwarz reduction.  

Note the general triangular feature of the equations, whereby certain generalised vielbeine and vectors appear more frequently than others.  More specifically, as one moves through equation \eqref{cDB1} to \eqref{cDB4}, as well as the generalised vielbeine and vectors that appeared before, a new generalised vielbein and vector contribute in turn.  This pattern is broken in equation \eqref{cDB4}, where $B_{\mu m_1\ldots m_7,n}$, which is associated with dual gravity degrees of freedom and the supersymmetry transformation of which gives generalised vielbein $e_{m_1 \ldots m_7, n AB}$ does not contribute. This is a completely general feature of the eleven-dimensional theory and, therefore, applies to any compactification.  An important consequence of this seems to be that any four-dimensional gauged theory obtained as a consistent reduction of $D=11$ supergravity cannot have gauge vectors associated with the gauging of these particular seven vectors.  This implies an additional constraint on the embedding tensor of any theory that is obtained from a reduction of $D=11$ supergravity.  However, we know that one can take a full set of 28 magnetic vectors in four dimensions and gauge these to obtain an SO(8) gauged maximal supergravity \cite{DIT}.  While it is true \cite{DIT} (see also \cite{dWN13}) that this theory is equivalent to the original SO(8) gauged maximal supergravity of \cite{dWNn8}, the very fact that a full set of magnetic vectors can be gauged in four dimensions and that this has no corresponding higher dimensional original is significant in understanding the extent to which the deformed SO(8) gauged maximal supergravities of \cite{DIT} can be realised as a reduction from $D=11$ supergravity.~\footnote{An interesting question is whether a deformation of the $D=11$ 56-bein $\cV$ \cite{GGN13} of the form
\begin{equation*}
\cV \longrightarrow \begin{pmatrix}
  \textup{e}^{i \omega} & 0 \\[2mm] 0 & \textup{e}^{-i \omega}
 \end{pmatrix}
\begin{pmatrix}
 \cV^{\tM \tN} \\[2mm] \cV_{\tM \tN}
\end{pmatrix},
\end{equation*}
in analogy with the rotation introduced in Ref.~\cite{dWN13}, allows the possibility of further gauging of magnetic vectors.  This would clearly point to 
the existence of a genuine deformation of $D=11$ supergravity.  Such a consistent 
deformation could then provide a higher dimensional origin of the deformed 
maximal SO(8) gauged supergravities of Ref.~\cite{DIT}.}

Let us first consider the connection coefficients $\cQ_{\mu}$ and $\cP_{\mu}$.  The $y$-dependence in both connection coefficients come from the same three terms, \textit{viz.}
\begin{align}
 \left[ {e^m}_a \partial_{m} B_{\mu}{}^{n} e_{n b} - ({e^p}_{a} \cD_{\mu} e_{p\, b}) \right], \qquad
{e_{\mu}}{}^{\alpha} F_{\alpha abc}, \qquad
e_{\mu \, \alpha} \eta^{\alpha \beta \gamma \delta} F_{\beta \gamma \delta a}.
\end{align}
Using the ansatz for $B_{\mu}{}^{m}$ and ${e^m}_a$, equations \eqref{ansBvec} and \eqref{anssieb} and property \eqref{dU} satisfied by $U$, it is simple to show that
\begin{equation}
 {e^m}_a \partial_{m} B_{\mu}{}^{n} e_{n b} - ({e^p}_{a} \cD_{\mu} e_{p\, b}) = - {{\hat{e}^p}}{}_{a} \left(\partial_{\mu} \hat{e}_{p\, b} - {f^n}_{pq} {\hat{B}_\mu}^q \hat{e}_{n \, b}\right).
\end{equation}
Hence, the $y$-dependence drops out.
Now, consider
\begin{equation}
 {e_{\mu}}{}^{\alpha} F_{\alpha abc} = (F_{\mu npq} - {B_\mu}^r F_{rnpq}) {{{e}^n}}{}_{a} {{{e}^p}}{}_{b} {{{e}^q}}{}_{c}.
\end{equation}
Notice that curved 7d indices enter only as dummy indices.  Furthermore, from equation \eqref{da3} we note that the $y$-dependence of the field strength in the second term is cancelled by the $y$-dependence of ${B_\mu}^r$ and the inverse siebenbein.  Therefore, the only potential obstacle to the dropping out of the $y$-dependence in the expression above is when a 7d derivative acts on the potential.  However, the 7d derivative always acts as an exterior derivative.  Hence, using equation \eqref{dU} and \eqref{da3}, we will always obtain a $y$-independent piece along with the appropriate $U$ contractions.  However, these $U$ factors will be cancelled for the same reason as stated above: that there is no free curved 7d index.  The same argument can be used to show the $y$-independence of the third term.  Therefore, we conclude that the connection coefficients are $y$-independent.

The eleven-dimensional fields enter the generalised vielbein postulates via the four generalised vielbeine and three of the vectors.  The reduction ans\"atze for the generalised vielbeine can be found using the ans\"atze for the fields that define them, equations \eqref{gv1}--\eqref{gv4}.  They are as follows:
\begin{align}
e^{m}_{AB} &= U^{-1/2} {(U^{-1})_n}^m \hat{e}^n_{AB}(x), \label{vb1} \\[3mm]
e_{mn AB} &= U^{-1/2} {U_{m}}^p {U_{n}}^q \hat{e}_{pq AB}(x) - a_{mnp} e^{p}_{AB}, \label{vb2} \\[3mm]
e_{m_1 \ldots m_5 AB} &= U^{-1/2} {U_{m_1}}^{n_1} \dots {U_{m_5}}^{n_5} \hat{e}_{n_1 \ldots n_5 AB}(x) - \frac{\sqrt{2}}{2} a_{[m_1 m_2 m_3} e_{m_4 m_5] AB} \notag \\[2mm]
&\hspace{50mm} - \left(a_{pm_1 \ldots m_5} + \frac{\sqrt{2}}{4} a_{p[m_1 m_2} a_{m_3 m_4 m_5]} \right) e^p_{AB}, \label{vb3} \\[3mm]
e_{m_1 \ldots m_7, n AB} &= U^{1/2} {U_{n}}^{p} \hat{e}_{m_1 \ldots m_7, p AB}(x) - a_{n[m_1 m_2} e_{m_3 \ldots m_7] AB} \notag \\[2mm] &\hspace{30mm} + \left( a_{n[m_1 \ldots m_5} - \frac{\sqrt{2}}{4} a_{n[m_1 m_2} a_{m_3 m_4 m_5} \right) e_{m_6 m_7] AB} \notag \\[2mm]
&\hspace{45mm} + \left( a_{n[m_1 \ldots m_5} - \frac{\sqrt{2}}{12} a_{n[m_1 m_2} a_{m_3 m_4 m_5} \right) a_{m_6 m_7]p} e^p_{AB}, \label{vb4}
\end{align}
where $\hat{e}^n_{AB}$, $\hat{e}_{pq AB}$, $\hat{e}_{n_1 \ldots n_5 AB}$ and $\hat{e}_{m_1 \ldots m_7, p AB}$ are the generalised vielbeine that appear in the torus reduction and are therefore directly related to the four-dimensional scalars.

The reduction ans\"atze for the vectors are found by using the fact that the supersymmetry transformation of the vectors \cite{GGN13},
\begin{align} \label{11dsusy1}
 \delta B_{\mu}{}^{m} &= \frac{\sqrt{2}}{8} \ e^{m}_{AB} \left[ 2
\sqrt{2} \overline{\varepsilon}^{A} \varphi_{\mu}^{B} +
\overline{\varepsilon}_{C} \gamma'_{\mu} \chi^{ABC} \right] \, +\,
\textup{h.c.}, \\[2mm] \label{11dsusy2}
  \delta B_{\mu m n} &= \frac{\sqrt{2}}{8} \ e_{mn AB} \left[ 2 \sqrt{2}
\overline{\varepsilon}^{A} \varphi_{\mu}^{B} +
\overline{\varepsilon}_{C} \gamma'_{\mu} \chi^{ABC} \right] \, +
\,\textup{h.c.}, \\[2mm] \label{11dsusy3}
   \delta B_{\mu m_1 \dots m_5} &= \frac{\sqrt{2}}{8} \ e_{m_1 \dots m_5
AB} \left[ 2 \sqrt{2} \overline{\varepsilon}^{A} \varphi_{\mu}^{B} +
\overline{\varepsilon}_{C} \gamma'_{\mu} \chi^{ABC} \right]  \, + \,
\textup{h.c.},\\[2mm]
    \delta B_{\mu m_1 \dots m_7, n} &= \frac{\sqrt{2}}{8} \ e_{m_1 \dots
m_7, n AB} \left[ 2 \sqrt{2} \overline{\varepsilon}^{A}
\varphi_{\mu}^{B} + \overline{\varepsilon}_{C} \gamma'_{\mu} \chi^{ABC}
\right] \, + \,  \textup{h.c.}, \label{11dsusy4}
\end{align}
should reproduce the respective generalised vielbeine.~\footnote{The factor of $U^{-1/2}$ are absent in the ans\"atze for the vectors because they are cancelled by a redefinition of the vierbein that contracts the fermions.}  The reduction ansatz for ${B_\mu}^m$ is give in equation \eqref{ansBvec}, while the reduction ansatz for $B_{\mu mn}$ and $B_{\mu m_1 \ldots m_5}$ are listed below:
\begin{align}
 B_{\mu mn} &= {U_{m}}^p {U_n}^q \hat{B}_{\mu pq}(x) - (U^{-1})_{p}{}^{q} {\hat{B}_\mu}{}^p a_{qmn}, \\[2mm]
B_{\mu m_1 \ldots m_5} &= {U_{m_1}}^{n_1} \dots {U_{m_5}}^{n_5} \hat{B}_{\mu n_1 \ldots n_5}(x) - \frac{\sqrt{2}}{2} a_{[m_1 m_2 m_3} \hat{B}_{|\mu| m_4 m_5]}(x) \notag \\[2mm]
&\hspace{51mm} - \left( a_{pm_1 \ldots m_5} - \frac{\sqrt{2}}{4} a_{p[m_1 m_2} a_{m_3 m_4 m_5]} \right) (U^{-1})_{q}{}^{p} {\hat{B}_\mu}{}^q.
\end{align}

Substituting the above ans\"atze into the generalised vielbein postulates \eqref{cDB1}--\eqref{cDB4}, a straightforward yet tedious calculation shows that the $y$-dependence in all the equations factorises out.  Importantly, we find that the two terms that vanished due to properties of Killing spinors on $S^7$ in the case of the $S^7$ compactification \cite{KKdual}, i.e.
\begin{equation*}
 \partial_{m}{B_\mu}^m \quad \text{and} \quad \partial_{[m} B_{|\mu|np]},
\end{equation*}
do not vanish in this case.  In particular,
\begin{align}
 \partial_{m}{B_\mu}^m &= \partial_m (U^{-1})_n{}^m {\hat{B}_\mu}{}^n = (U^{-1})_n{}^m \partial_{m} \text{log}U {\hat{B}_\mu}{}^n, \\[2mm]
 \partial_{[m} B_{|\mu|np]} &= {U_{[m}}^q {U_n}^r {U_{p]}}^s {f^t}_{qr} \hat{B}_{\mu st} - \partial_{[m}\left(a_{np]q}(U^{-1})_r{}^q \right) {\hat{B}_\mu}{}^r,
\end{align}
where in the first line we have used equation \eqref{Uunimod}.

The generalised vielbein postulates reduce to the following equations
\begin{align} \label{rgvp1}
& \partial_{\mu} \hat{e}^{m}_{AB} - {f^{m}}_{pq} {\hat{B}_\mu}{}^p \hat{e}^q_{AB} + \cQ_{\mu}^{C}{}_{[A} \hat{e}^{m}_{B]C} + \cP_{\mu ABCD} \hat{e}^{mCD} = 0, \\[3mm] \label{rgvp2}
&\partial_\mu \hat{e}_{mnAB} -2 {f^p}_{q[m} \hat{e}_{n]p AB} {\hat{B}_\mu}{}^q + 3 {f^q}_{[mn} \hat{B}_{|\mu|p]q} \hat{e}^p_{AB} 
+ \frac{1}{6} g_{mnpq} {\hat{B}_\mu}{}^p \hat{e}^q_{AB} \notag \\[3mm]
&\hspace{80mm}+ \cQ_{\mu}^{C}{}_{[A} \hat{e}_{mnB]C} + \cP_{\mu ABCD} \hat{e}_{mn}{}^{CD} = 0, \\[3mm] \label{rgvp3}
&\partial_\mu \hat{e}_{m_1 \dots m_5 AB}  + 5 {f^p}_{q[m_1} \hat{e}_{m_2 \ldots m_5]p AB} {\hat{B}_\mu}{}^q -\frac{3\sqrt{2}}{2} {f^p}_{[m_1 m_2} \hat{B}_{|\mu p|m_3} \hat{e}_{m_4 m_5] AB} \notag \\[3mm]
&\hspace{10mm}+ 15 {f^p}_{[m_1 m_2} \hat{B}_{|\mu p| m_3 m_4 m_5 q]} \hat{e}^q_{AB} +\frac{\sqrt{2}}{12} {\hat{B}_\mu}{}^p g_{p[m_1 m_2 m_3} \hat{e}_{m_4 m_5] AB} +\frac{\sqrt{2}}{8} \hat{B}_{\mu [m_1 m_2} g_{m_3 m_4 m_5p]} \hat{e}^{p}_{AB}
\notag \\[3mm]
&\hspace{28mm} -\frac{1}{6!} \fr \eta_{pqm_1 \ldots m_5} {\hat{B}_\mu}{}^p \hat{e}^q_{AB}
+ \cQ_{\mu}^{C}{}_{[A} \hat{e}_{m_1 \dots m_5 B]C} + \cP_{\mu ABCD} \hat{e}_{m_1 \dots m_5}{}^{CD}=0, \\[15pt] \label{rgvp4}
& \partial_\mu \hat{e}_{m_1 \dots m_7, n AB} + {f^{p}}_{qn} {\hat{B}_\mu}{}^q \hat{e}_{m_1 \dots m_7, p AB}
-5 {f^p}_{[m_1 m_2} \hat{B}_{|\mu p|m_3} \hat{e}_{m_4 \ldots m_7]n AB} \notag \\[3mm]
&\hspace{10mm}+5 {f^p}_{[m_1 m_2} \hat{B}_{|\mu p|m_3 \ldots m_6} \hat{e}_{m_7]n AB}
+ \frac{5}{18} {\hat{B}_\mu}{}^p g_{p[m_1 m_2 m_3} \hat{e}_{m_4 \ldots m_7]n AB} 
+\frac{\sqrt{2}}{24} \hat{B}_{\mu [m_1 m_2} g_{m_3 \ldots m_6} \hat{e}_{m_7]n AB} \notag \\[3mm]
&\hspace{28mm} +\frac{1}{3\cdot 7!} \fr \eta_{m_1 \ldots m_7} {\hat{B}_\mu}{}^p \hat{e}_{pn AB}
+ \cQ_{\mu}^{C}{}_{[A} \hat{e}_{m_1 \dots m_7, n B]C} + \cP_{\mu ABCD} \hat{e}_{m_1 \dots m_7, n}{}^{CD} =0.
\end{align}
As emphasised before, the $y$-independent, hatted generalised vielbeine and vectors in the generalised vielbein postulates above are directly related to the respective four-dimensional quantities. In particular, since the reduction of these eleven-dimensional quantities is taken to be that of a simple toroidal nature, the conversion of `curved' SU(8) indices $A,B,C,\ldots$ to flat SU(8) indices $i,j,k,\ldots$ is trivial.

With this in mind, define an E$_{7(7)}$ vielbein~\footnote{Strictly speaking, $\cV$ is not 
an E$_{7(7}$ group element because it is acted upon by SU(8) transformations on the right,
whereas the indices on the left are to be regarded as SL(8) indices. The true E$_{7(7)}$ group 
element is obtained by a complex rotation of this matrix (see, for example, Ref.~\cite{BHN} for more details).}
\begin{equation}\label{56bein}
 \cV_{\cM \, ij}=\big(\cV_{\tM \tN\, ij} \,, \, \cV^{\tM \tN}{}_{ij} \big)
\end{equation}
that is related to the hatted generalised vielbeine via the following relations:
\begin{align}
 \cV^{m8}{}_{ij} &= \frac{\sqrt{2}}{8} i\, \hat{e}^m_{ij}, \hspace{33mm} \cV_{mn\,
ij} = - \frac{3}{2} i\, \hat{e}_{mn\, ij}, \notag \\[2mm]
\cV^{mn}{}_{ij} &= \frac{3}{2} i\, {\eta}^{mnp_1\dots p_5}
\hat{e}_{p_1\dots p_5\, ij}, \hspace{10mm}
\cV_{m8\,ij} = -\frac{9\sqrt{2}}{2} i\, {\eta}^{n_1\dots n_7}
\hat{e}_{n_1\dots n_7, m\, ij}. 
\label{Vdef}
\end{align}
As expected $ \cV$ satisfies the E$_{7(7)}$ properties, as can be checked explicitly using equations \eqref{gv1}--\eqref{gv4} and \eqref{Vdef},
\begin{align}
 \cV_{\cM ij} \cV_{\cN}{}^{ij} - \cV_{\cM}{}^{ij} \cV_{\cN ij} &= i\, \Omega_{\cM \cN}, \notag \\[2mm]
 \Omega^{\cM \cN} \cV_{\cM}{}^{ij} \cV_{\cN kl} &= i\, \delta^{ij}_{kl}, \notag \\[2mm]
 \Omega^{\cM \cN} \cV_{\cM}{}^{ij} \cV_{\cN}{}^{kl} &= 0, 
\end{align}
where the symplectic form $\Omega$ is such that
\begin{align}
 \Omega^{\tM \tN}{}_{\tP \tQ}&=\delta^{\tM \tN}_{\tP \tQ}, \qquad \Omega_{\tM \tN}{}^{\tP \tQ}=-\delta_{\tM \tN}^{\tP \tQ}, \notag \\
 \Omega_{\tM \tN\, \tP \tQ}&=0, \qquad \hspace{3mm} \Omega^{\tM \tN\, \tP \tQ}=0.
\end{align}

Similarly, we combine the vectors into a {\bf56} of E$_{7(7)}$ defined by
\begin{equation}
\cA_{\mu}^{\cM} = (\cA_{\mu}^{\tM \tN},\, \cA_{\mu \, \tM \tN})
\end{equation}
where
\begin{align}
  {\cA_{\mu}}^{m 8} &= -\frac{1}{2} {\hat{B}_{\mu}}^m, \hspace{36.5mm}
\cA_{\mu\, mn} = -3\sqrt{2}\, \hat{B}_{\mu mn}, \notag \\[2mm]
{\cA_{\mu}}^{m n} &= -3\sqrt{2}\, {\eta}^{mnp_1\dots
p_5} \hat{B}_{\mu p_1 \ldots p_5}, \hspace{9mm}
\cA_{\mu\, m8} = -18\, {\eta}^{n_1\dots n_7} \hat{B}_{\mu n_1
\dots n_7, m}. 
\label{Bdef}
\end{align}

In the notation introduced above, the supersymmetry transformations of the generalised vielbeine and vectors takes a very compact form
\begin{gather}
 \hspace{-5mm} \delta \cV_{\cM \, ij} = \sqrt{2} \Sigma_{ijkl} \cV_{\cM}{}^{kl}, \\[3mm]
 \delta \cA_{\mu}{}^{\cM} = i \, \Omega^{\cM \cN} \cV_{\cN \, ij} \left(2 \sqrt{2} \overline{\varepsilon}^{i}\varphi_{\mu}^{j} + \overline{\varepsilon}_{k} \hat{\gamma}_{\mu} \chi^{ijk} \right) \,+ \, \textup{h.c.}.
\end{gather}

In order to relate our results for the Scherk-Schwarz reduction with the four-dimensional understanding of gaugings as embodied in the embedding tensor formalism, we need to rewrite the reduced generalised vielbein postulates \eqref{rgvp1}--\eqref{rgvp4} in terms of the notation introduced above, that is in terms of E$_{7(7)}$ objects $\cV$ and $\cA$.  A straightforward calculation shows that upon substitution of $\cV$ and $\cA$ components, as defined by equation \eqref{Vdef} and \eqref{Bdef}, equations \eqref{rgvp1}--\eqref{rgvp4} become
\begin{align} \label{fgvp1}
& \partial_{\mu} \cV^{m}_{ij} + \cQ_{\mu}^{k}{}_{[i} \cV^{m}_{j]k} - \cP_{\mu\, ijkl} \cV^{mkl} + 2 {\cA_\mu}^p {f^{m}}_{pq} \cV^q_{ij}= 0, \\[3mm] \label{fgvp2}
&\partial_\mu \cV_{mn\, ij} + \cQ_{\mu}^{k}{}_{[i} \cV_{|mn|j]k} - \cP_{\mu\, ijkl} \cV_{mn}{}^{kl} \notag \\[2mm]
& \hspace{40mm} +4  {\cA_\mu}^p \delta_{[m}^{[r} {f^{s]}}_{n]p}  \cV_{rs ij} +6 \cA_{\mu\, pq} \delta^{[p}_{[r} {f^{q]}}_{mn]} \cV^r_{ij} + 2 \sqrt{2} \,  {\cA_\mu}^p g_{mnpq} \cV^q_{ij} = 0, \\[3mm] \label{fgvp3}
&\partial_\mu \cV^{mn}_{ij} + \cQ_{\mu}^{k}{}_{[i} \cV^{mn}_{j]k} - \cP_{\mu\, ijkl} \cV^{mn\, kl}
-4 {\cA_\mu}^p \delta^{[m}_{[r} {f^{n]}}_{s]p} \cV^{rs}_{ij} +\frac{1}{2}  \cA_{\mu \, pq} \eta^{mnturs[p} {f^{q]}}_{tu}\cV_{rs\, ij} \notag \\[3mm]
&\hspace{35mm} -2\cA_{\mu}{}^{pq} \delta_{r}^{[m} {f^{n]}}_{pq} \cV^r_{ij} +\frac{\sqrt{2}}{6} {\cA_\mu}^p \eta^{mnqrstu} g_{pqrs} \cV_{tu\, ij}  \notag \\[3mm]
& \hspace{50mm}-\frac{\sqrt{2}}{12} \cA_{\mu pq} \delta^{[m}_{s} \eta^{n]pqr_1 \ldots r_4} g_{r_1 \ldots r_4} \cV^{s}_{ij} +4\sqrt{2} \fr {\cA_\mu}^p \delta^{mn}_{pq} \cV^q_{ij}=0, \\[15pt] \label{fgvp4}
& \partial_\mu \cV_{m\, ij} + \cQ_{\mu}^{k}{}_{[i} \cV_{m\, j]k} - \cP_{\mu\, ijkl} \cV_{m}{}^{kl}
-2 {\cA_\mu}^p {f^{q}}_{pm} \cV_{q\, ij}
+3 \cA_{\mu\, pq} \delta^{[p}_{[m} {f^{q]}}_{rs]}  \cV^{rs}_{ij} + \cA_{\mu}{}^{pq} \delta^{[r}_{m} {f^{s]}}_{pq} \cV_{rs\, ij} \notag \\[3mm]
&\hspace{10mm}
+ \sqrt{2}\, {\cA_\mu}^p g_{pqrm} \cV^{qr}_{ij} 
-\frac{\sqrt{2}}{24} \cA_{\mu pq} \eta^{pqr_1 \ldots r_4[s}\delta^{t]}_{m} g_{r_1 \ldots r_4} \cV_{st\, ij} - 2\sqrt{2}\, \fr {\cA_\mu}^p \delta^{rs}_{pm}  \cV_{rs\, ij} =0.
\end{align}

Now, the components of $X_\cM$ in terms of GL(7) indices can be simply read off by comparing equation \eqref{embedeqn} and equations \eqref{fgvp1}--\eqref{fgvp4} listed above~\footnote{For brevity, we have left out a factor of the gauge coupling $g$ in these expressions.}
\begin{align}
 X_{m8}{\,}^{p8}{\,}_{r8} &= - X_{m8}{\,}_{r8}{\,}^{p8} = -\frac{1}{2} f^{p}{}_{mr}, \hspace{31.1mm} X_{m8}{\,}^{pq}{\,}_{r8} = - X_{m8}{\,}_{r8}{\,}^{pq} = - \sqrt{2} \delta^{pq}_{mr} \fr, \notag \\[5pt]
X_{m8}{\,}^{pq}{\,}_{rs} &=  -  X_{m8}{\,}_{rs}{\,}^{pq} = 2 \delta^{[p}_{[r} f^{q]}{}_{s]m}, \hspace{28.9mm} X_{mn}{\,}^{pq}{\,}_{r8} = - X_{mn}{\,}_{r8}{\,}^{pq} =  \delta^{[p}_{r} f^{q]}{}_{mn}, \notag \\[5pt]
 X^{mn}{\,}_{p8}{\,}_{rs} &= \;\, X^{mn}{\,}_{rs}{\,}_{p8} \; = - 3 \delta^{[m}_{[p} f^{n]}{}_{rs]}, \hspace{25.1mm} 
 X^{mn}{\,}^{pq}{\,}^{rs} = - \frac{1}{2} \eta^{pqrstu[m} f^{n]}{}_{tu}, \notag \\[5pt]
 X^{mn}{\,}_{p8}{\,}^{rs} &= - X^{mn}{\,}^{rs}{\,}_{p8} = - \frac{\sqrt{2}}{24} \delta^{[r}_{p} \eta^{s]mntuvw} g_{tuvw}, \hspace{7.3mm} X_{m8}{\,}^{pq}{\,}^{rs} = - \frac{\sqrt{2}}{12} \eta^{pqrstuv} g_{mtuv}, \notag \\[5pt] 
 X_{m8}{\,}_{p8}{\,}_{rs} &= \;\, X_{m8}{\,}_{rs}{\,}_{p8} \;\, =  - \frac{\sqrt{2}}{2} g_{mprs}.  
  \label{X2}
\end{align}
The components of $X_{\cM}$ presented above agree in their general form with the components given already in the literature \cite{ssdft}.~\footnote{There are some discrepancies 
in numerical factors (see equation (4.16) of Ref.~\cite{ssdft}).
In any case, here we verify that both the linear and quadratic constraints are satisfied for the components of $X_{\cM}$ given in equation \eqref{X2}.}  
Written in terms of SL(8) indices, they take the form of the general solution given in equations \eqref{X} with
\begin{align} \label{tAcompts}
 \tA_{1\, 88} = -\frac{8\sqrt{2}}{3} \fr, \qquad {\tA_2}^{m}{}_{np8} = -\frac{8}{3} {f^m}_{np}, \qquad \tA_{2 \, 8}{}^{mnp}=\frac{\sqrt{2}}{9} \eta^{mnpr_1\ldots r_4} g_{r_1\ldots r_4},
\end{align}
and all other components vanishing.  The appearance of these structures can be understood from a group-theoretic point of view by considering the branching of the {\bf 912} representation of E$_{7(7)}$ in which the embedding tensor lives with respect to GL(7) \cite{Mtwistedtorus, ssdft, SamLect}
\begin{align}
\bf 912 \rightarrow 1_{+7} + \overline{35}_{+5} + (7 + 140)_{+3} &+ (\bf \overline{21} + \overline{28} + \overline{224})_{+1} \notag \\
&+\bf ( 21 + 28 + 224)_{-1} + (\overline{7}+\overline{140})_{-3} + 35_{-5} + 1_{-7},
\end{align}
where the subscript represents the charge under GL(1) $\subset$ GL(7).  Hence \cite{Mtwistedtorus, ssdft}
\begin{align}
 \fr &\longleftrightarrow {\bf 1_{+7}} \notag \\
 g_{mnpq} &\longleftrightarrow {\bf \overline{35}_{+5}} \notag \\
 {f^p}_{mn} &\longleftrightarrow {\bf 140_{+3}} \notag \\
 {f^p}_{pm} &\longleftrightarrow {\bf 7_{+3}}. \notag
\end{align}
Of course, ${f^p}_{pm}=0$, so ${\bf 7_{+3}}$ does not contribute.

Note that we have used
\begin{equation} \label{eta8}
 \eta_{m_1\ldots m_7 8} = \eta_{m_1\ldots m_7}.
\end{equation}

The quadratic constraint \eqref{cons:quad} is satisfied for the $X_{\cM}$ derived from the generalised vielbein postulates. The constraints must be verified for each component and they are shown to be satisfied using Schouten identities, the unimodularity property \eqref{funi}, the Jacobi identity \eqref{ffjac} and the background Bianchi identity \eqref{fgjac}. We refer the reader to appendix \ref{app:quad} for details.

The calculations involved in the verification of the quadratic constraint are highly non-trivial. However, the fact that $X_{\cM}$ as derived from the \emph{eleven-dimensional} generalised vielbein postulates not only satisfy the linear constraint, but also the more non-trivial quadratic constraint shows that there is indeed a {\textit{bona fide}} gauge algebra for the gauging in the reduction.  More generally, it points yet again to the deep relation between our eleven-dimensional formalism, developed in Refs.~\cite{GGN13, KKdual}, and the embedding tensor formalism \cite{NSmaximal3,NScomgauge3,dWSTlag,dWSTgauge, dWSTmax4} that describes gauged supergravity. 

Note that the verification of the linear and quadratic constraints did not require the use of the background consistency equations \eqref{bkg:einstein} and \eqref{bkg:maxwell}.  These are extra constraints that must be satisfied by the background solution if the reduction is to be consistent.

\section{Scherk-Schwarz reduction with no flux} \label{sec:flat}

An object $\Theta_{\cM}{}^{\alpha}$, satisfying the embedding tensor constraints is guaranteed to have at most half-maximal row-rank \cite{SamLect} as was explained in section \ref{sec:embedding}.  However, even though we have shown that $\Theta_{\cM}{}^{\alpha}$ as derived from the  generalised vielbein postulates satisfies the embedding tensor constraints, it is not 
immediately obvious that always less than 28 vectors will be gauged, as is required by consistency.  In fact a naive counting suggests that 49 vectors contribute, since this is the number of vectors that remain in the generalised vielbein postulates after the reduction ans\"atze are substituted in.  This is in contrast to the case of the $S^7$ reduction considered in \cite{KKdual}.  There it is clear from the onset that $B_{\mu\, mn}$ drop out of the generalised vielbein postulates because of properties of Killing vectors.  This leaves $A_{\mu}{}^{m}$ and $A_{\mu}{}^{mn}$, which are indeed the 28 vectors that are gauged in the $S^7$ reduction.  

The fact that general results of the embedding tensor formalism guarantee that less than or equal to 28 vectors are gauged means that our naive counting of the contributing vectors is over-simplified and that constraints such as those placed on structure constants ${f^{p}}_{mn}$ for consistency of the reduction will conspire to reduce the number of gauged vectors to less than 28.

In this section, we explicitly demonstrate this for the simplifying case corresponding to the original reduction considered in \cite{scherkschwarz}, where there is no flux, i.e.
\begin{equation}
 \fr=0, \qquad g_{mnpq}=0.
\end{equation}
The background equation \eqref{bkg:4d} implies that the four dimensional spacetime is Minkowski and that the group under consideration is ``flat'' \cite{scherkschwarz}, i.e.
\begin{equation}
2 \delta_{p q} \delta^{r s} f^{p}{}_{m r} f^{q}{}_{n s} + 2 f^{p}{}_{m q} f^{q}{}_{n p} - \delta_{m p} \delta_{n q} \delta^{rs} \delta^{tu} f^{p}{}_{r t} f^{q}{}_{s u} = 0.
\end{equation}
In this case the generalised vielbein postulates \eqref{rgvp1}--\eqref{rgvp4} take a simpler form
\begin{align} \label{nfrgvp1}
& \partial_{\mu} \hat{e}^{m}_{AB} - {f^{m}}_{pq} {\hat{B}_\mu}{}^p \hat{e}^q_{AB} + \cQ_{\mu}^{C}{}_{[A} \hat{e}^{m}_{B]C} + \cP_{\mu ABCD} \hat{e}^{mCD} = 0, \\[3mm] \label{nfrgvp2}
&\partial_\mu \hat{e}_{mnAB} -2 {f^p}_{q[m} \hat{e}_{n]p AB} {\hat{B}_\mu}{}^q + 3 {f^q}_{[mn} \hat{B}_{|\mu|p]q} \hat{e}^p_{AB} + \cQ_{\mu}^{C}{}_{[A} \hat{e}_{mnB]C} + \cP_{\mu ABCD} \hat{e}_{mn}{}^{CD} = 0, \\[3mm] \label{nfrgvp3}
&\partial_\mu \hat{e}_{m_1 \dots m_5 AB}  + 5 {\hat{B}_\mu}{}^q {f^p}_{q[m_1} \hat{e}_{m_2 \ldots m_5]p AB} -\frac{3\sqrt{2}}{2} {f^p}_{[m_1 m_2} \hat{B}_{|\mu p|m_3} \hat{e}_{m_4 m_5] AB} \notag \\[3mm]
&\hspace{10mm}+ 15 {f^p}_{[m_1 m_2} \hat{B}_{|\mu p| m_3 m_4 m_5 q]} \hat{e}^q_{AB} + \cQ_{\mu}^{C}{}_{[A} \hat{e}_{m_1 \dots m_5 B]C} + \cP_{\mu ABCD} \hat{e}_{m_1 \dots m_5}{}^{CD}=0, \\[15pt] \label{nfrgvp4}
& \partial_\mu \hat{e}_{m_1 \dots m_7, n AB} + {f^{p}}_{qn} {\hat{B}_\mu}{}^q \hat{e}_{m_1 \dots m_7, p AB}
-5 {f^p}_{[m_1 m_2} \hat{B}_{|\mu p|m_3} \hat{e}_{m_4 \ldots m_7]n AB} \notag \\[3mm]
&\hspace{10mm}+5 {f^p}_{[m_1 m_2} \hat{B}_{|\mu p|m_3 \ldots m_6} \hat{e}_{m_7]n AB} + \cQ_{\mu}^{C}{}_{[A} \hat{e}_{m_1 \dots m_7, n B]C} + \cP_{\mu ABCD} \hat{e}_{m_1 \dots m_7, n}{}^{CD} =0.
\end{align}
A simple example of a flat group is given by \cite{scherkschwarz}
\begin{equation}
 {U_{m}}{}^n= (\textup{exp}M y^1)_m{}^{n},
\end{equation}
where the seven-dimensional coordinates $y^m=(y^1,y^{\tilde{m}})$ with $\tilde{m}=2,\ldots,7$ and $M$ is a constant traceless matrix with zeros in the first row and column, i.e.
\begin{equation}
M_{m}{}^{n} = \begin{pmatrix}
              \ 0\ & \underline{0}^\textup{T} \\[3mm]\ \underline{0}\ & \tilde{M}_{\tilde{m}}{}^{\tilde{n}}
              \end{pmatrix}.
\end{equation}
Using the fact that
\begin{equation}
 \partial_m {U_{n}}{}^p = \delta_m^1 U_n{}^{q} M_{q}{}^p,
\end{equation}
we find that
\begin{equation}
 {f^{p}}_{mn} = 2 M_{[m}{}^p \delta_{n]}^1.
\end{equation}
In particular, we find that the only non-zero components of the structure constant are ${f^{\tilde{p}}}_{1\tilde{n}}$.  Inspecting the generalised vielbein postulates \eqref{nfrgvp1}--\eqref{nfrgvp4} we find that $\hat{B}_{\mu\, mn}$ and $\hat{B}_{\mu\, m_1 \ldots m_5}$ enter the equations in the form
\begin{equation*}
{f^{q}}_{[mn} \hat{B}_{\mu\, p] q} \qquad \textup{and} \quad {f^{p}}_{[m_1 m_2} \hat{B}_{\mu\, m_3 \ldots m_6] p}.
\end{equation*}
Hence, only
\begin{equation*}
 \hat{B}_{\mu \tilde{m} \tilde{n}} \qquad \textup{and} \quad \hat{B}_{\mu \tilde{m}_1 \ldots \tilde{m}_5}
\end{equation*}
contribute.  Along with $\hat{B}_{\mu}{}^{1}$ and $\hat{B}_{\mu}{}^{\tilde{m}}$ this gives a total of
\begin{equation*}
28=1+6+6+15 = 13\ \textup{electric} + 15\ \textup{magnetic}
\end{equation*}
vectors appearing in the generalised vielbein postulates, which is kinematically consistent. 
Of course, one should here distinguish between the kinematics of the gauge couplings
and the dynamics of the theory, which determines the vacuum and thus decides which vectors will remain as massless
gauge bosons, and which will acquire a mass through spontaneous symmetry breaking.
Indeed, for generic Scherk-Schwarz compactifications,  the majority of the candidate 
28 vectors fields will become massive in the reduction and can therefore not be gauged.  In fact, $\hat{B}_{\mu}{}^{1}$ is the only vector that becomes gauged in the reduced theory.  An analysis of all possible gaugings from a Scherk-Schwarz reduction with no background flux is given 
in Ref.~\cite{ADFL}. It is shown that only electric vectors become gauged in this case.

In general, the Scherk-Schwarz reduction with background fluxes will have less than or equal to 28 gauge vectors contributing, kinematically, as is expected from general arguments.  However, the distribution between electric and magnetic vectors can be varied---although as pointed out before, no more than 21 magnetic vectors can be gauged in this symplectic frame.  In the context of Scherk-Schwarz flux compactifications this has already been observed in \cite{Mtwistedtorus}.

\section{Concluding remarks} \label{sec:con}

In this paper, we have investigated the Scherk-Schwarz reduction of $D=11$ supergravity with background flux.  In this case, the reduction ansatz immediately gives a relation between the 56-bein in eleven dimensions and the 56-bein that parametrises the scalars in four dimensions, equations \eqref{vb1}--\eqref{vb4}.  In this form, the reduction ansatz is applied to the generalised vielbein postulates yielding the embedding tensor of the respective gauged maximal theories in four dimensions.  Furthermore, the reduction ansatz written in the form \eqref{vb1}--\eqref{vb4} is suggestive of the fact that Scherk-Schwarz flux reductions can be thought of as an E$_{7(7)}$ generalised Scherk-Schwarz reduction of the form
\begin{align} \label{gssa}
 \cV_{\cM \, AB}(x,y) &= \cU_{\cM}{}^{\cN}(y)\ \hat{\cV}_{\cN \, AB}(x), \\[2mm]
 \cB_{\mu \, \cM}(x,y) &= U^{1/2}\, \cU_{\cM}{}^{\cN}(y)\ \cA_{\mu \, \cN}(x), \label{gssb}
\end{align}
where 
{\small \begin{equation}
 \cV_{\cM \, AB} = \begin{pmatrix}
  \cV_{m8\, AB} \\[2mm] \cV^{mn}{}_{AB} \\[2mm] \cV_{mn\, AB} \\[2mm] \cV^{m8}{}_{AB}
 \end{pmatrix}, \qquad 
\cB_{\mu \, \cM} = \begin{pmatrix}
                    \cB_{\mu \, m8} \\[2mm] \cB_{\mu}{}^{mn} \\[2mm] \cB_{\mu \,mn} \\[2mm] \cB_{\mu}{}^{m8}
                   \end{pmatrix},
\end{equation}}

\noindent and $\hat{\cV}_{\cM \, AB}$ and $\cA_{\mu \, \cN}$ (similarly defined) are the 56-bein and the set of 56 vectors appropriate for the torus reduction, respectively.  Moreover, $\cU(y)$ is an E$_{7(7)}$ matrix of the form
{ \small
\begin{equation}\label{cU} 
 \begin{pmatrix} 
    U^{1/2} U_{m}{}^{p} & 3\sqrt{2} U^{1/2} a_{mrs} (U^{-1})_{p}{}^{r} (U^{-1})_{q}{}^{s} & U^{-1/2} {S_{+}^{rs}}_{m} U_{r}{}^{p} U_{s}{}^{q} & U^{-1/2} S_{m\, s} (U^{-1})_{p}{}^{s} \\[3mm] 
    0 & U^{1/2} (U^{-1})_{p}{}^{m} (U^{-1})_{q}{}^{n} & U^{-1/2} S^{mn\, rs} U_{r}{}^{p} U_{s}{}^{q}  & -2 U^{-1/2} {S_{-}^{mn}}_{s} (U^{-1})_{p}{}^{s} \\[3mm] 
    0 & 0 & U^{-1/2} U_{m}{}^{p} U_{n}{}^{q} & 6\sqrt{2} U^{-1/2} a_{mnr} (U^{-1})_{p}{}^{r} \\[3mm] 
    0 & 0 & 0 &  U^{-1/2} (U^{-1})_{p}{}^{m}               
   \end{pmatrix},
\end{equation}}

\noindent where
\begin{align}
 {S_{\pm}^{mn}}_{s} &=3\sqrt{2} \eta^{mnr_1\dots r_5} \left(a_{s r_1 \dots r_5} \pm \textstyle{\frac{\sqrt{2}}{4}} a_{s r_1 r_2} a_{r_3 r_4 r_5}\right), \\
 S_{m\, n} &= -36 \eta^{r_1 \dots r_7} a_{m r_1 r_2} \left( a_{n r_3 \dots r_7} - \textstyle{\frac{\sqrt{2}}{12}} a_{n r_3 r_4} a_{r_5 r_6 r_7} \right), \\
 S^{mn \, pq} &= \textstyle{\frac{\sqrt{2}}{2}} \eta^{mnpqr_1 r_2 r_3 } a_{r_1 r_2 r_3}.
\end{align}

Equation \eqref{gssa} is to compared with equation (64) of Ref.~\cite{GGN13}:
\begin{equation}
 \cV_{\cM \, AB}(x,y) = \cV_{\cM}{}^{\cA}(x,y)\ \Gamma_{\cA \, AB},
\end{equation}
where 
{\small \begin{equation}
\Gamma_{\cA \, AB} = \begin{pmatrix}
  \Gamma_{a\, AB} \\[2mm] \Gamma^{ab}_{AB} \\[2mm] i \Gamma_{ab\, AB} \\[2mm] i \Gamma^{a}_{AB}
 \end{pmatrix}.
\end{equation}}

\noindent In this case, one finds that the form of matrix $\cU(y)$ is exactly the same as the form of $\cV_{\cM}{}^{\cA}$ with the following identifications
\begin{equation}
 U_{m}{}^{n} \longleftrightarrow e_{m}{}^{a}, \qquad a_{mnp} \longleftrightarrow A_{mnp}, \qquad a_{m_1 \dots m_6} \longleftrightarrow A_{m_1 \dots m_6}.
\end{equation}
In particular, in Ref.~\cite{GGN13}, $\cV_{\cM}{}^{\cA}$ is identified with the E$_{7(7)}$ 
coset element constructed in Ref.~\cite{hillmann}.

An interesting question is whether new reductions can be found by considering an 
ansatz of the form \eqref{gssa}, \eqref{gssb}. A direction related to this is pursued 
in \cite{Berman:2012uy, Musaev:2013rq, Aldazabal:2013mya} in the context of 
extended generalised geometry, where $\cU_{\cM}{}^{\cN}$ is assumed to depend 
on all extended coordinates. One should, however, keep in mind that \eqref{cU}
is already the most general E$_{7(7)}$ matrix (albeit in a triangular gauge), which
does not leave much room for more exotic possibilities.

\vspace{0.7cm}\noindent
{\bf Acknowledgments:} We are grateful to Bernard de Wit and Henning Samtleben for
stimulating discussions. H.G.~and M.G.~would like to thank the Max-Planck-Institut f\"{u}r Gravitationsphysik (AEI) and ENS, Lyon and in particular H.N.~and Henning Samtleben for their generous hospitality while this project was being carried out.  H.G.~and M.G.~are supported 
by King's College, Cambridge. H.N.~was supported by the Gay-Lussac-Humboldt Prize during his stay at ENS, Lyon where this work was completed.

\newpage
\appendix

\section{E$_{7(7)}$ algebra and identities} \label{app:e7}

In this appendix we review the SL(8) decomposition of the E$_{7(7)}$ algebra. In such a decomposition, the generators in the adjoint representation can be written
\begin{align}
 (t^{\tM}{}_{\tN})^{\tP\tQ}{}_{\tR\tS} &= 2 \left(\delta^{\tP\tQ}_{\tN[\tS} \delta^{\tM}_{\tR]} - \frac{1}{8} \delta^{\tM}_{\tN} \delta^{\tP\tQ}_{\tR\tS} \right), \qquad
(t^{\tM}{}_{\tN})_{\tR\tS}{}^{\tP\tQ} = - 2 \left(\delta^{\tP\tQ}_{\tN[\tS} \delta^{\tM}_{\tR]} - \frac{1}{8} \delta^{\tM}_{\tN} \delta^{\tP\tQ}_{\tR\tS} \right),\\
(t_{\tP\tQ\tR\tS})^{\tT_1 \ldots \tT_4} &= \delta^{\tT_1 \ldots \tT_4}_{\tP\tQ\tR\tS}, \hspace{35mm}  (t_{\tP\tQ\tR\tS})_{\tT_1 \ldots \tT_4} = \frac{1}{4!} \eta_{\tP\tQ\tR\tS \tT_1 \ldots \tT_4}.
\end{align}
It can be explicitly checked that the generators satisfy the following familiar commutation relations
\begin{gather}
 \left[ t^{\tM}{}_{\tN}, t^{\tP}{}_{\tQ} \right] = \delta^{\tM}_{\tQ} t^{\tP}{}_{\tN} - \delta^{\tP}_{\tN} t^{\tM}{}_{\tQ}, \qquad
\left[ t^{\tM}{}_{\tN}, t_{\tP\tQ\tR\tS} \right] = - 4 \left( \delta_{[\tP}^{\tM} t_{\tQ\tR\tS]\tN} + \frac{1}{8} \delta^{\tM}_{\tN} t_{\tP\tQ\tR\tS} \right), \\
  \left[ t_{\tM\tN\tP\tQ}, t_{\tR\tS\tT\tU} \right] = \frac{1}{72} \left( \eta_{\tV\tM\tN\tP\tQ[\tR\tS\tT} t^{\tV}{}_{\tU]} - \eta_{\tV\tR\tS\tT\tU[\tM\tN\tP} t^{\tV}{}_{\tQ]} \right).
\end{gather}
It is sometimes convenient to also define coset generators with upper indices
\begin{equation}
t^{\tM\tN\tP\tQ} = \frac1{4!} \eta^{\tM\tN\tP\tQ\tR\tS\tT\tU} t_{\tR\tS\tT\tU}
\end{equation}
keeping in mind that these are not independent generators.
Furthermore, the components of the Killing metric are 
\begin{align}
 \kappa^{\tM}{}_{\tN},{}^{\tP}{}_{\tQ} &= 12 \left(\delta^{\tM}_{\tQ} \delta^{\tP}_{\tN} - \frac{1}{8} \delta^{\tM}_{\tN} \delta^{\tP}_{\tQ} \right), \hspace{17mm} \kappa_{\tM\tN\tP\tQ,\tR\tS\tT\tU} = \frac{2}{4!} \eta_{\tM\tN\tP\tQ\tR\tS\tT\tU}, \notag \\
 (\kappa^{-1})_{\tN}{}^{\tM},{}_{\tQ}{}^{\tP} &= \frac{1}{12} \left(\delta^{\tM}_{\tQ} \delta^{\tP}_{\tN} - \frac{1}{8} \delta^{\tM}_{\tN} \delta^{\tP}_{\tQ} \right),\hspace{10mm} (\kappa^{-1})^{\tM\tN\tP\tQ,\tR\tS\tT\tU} = \frac{1}{2\cdot 4!} \eta^{\tM\tN\tP\tQ\tR\tS\tT\tU}.
\end{align}

\section{The quadratic constraint} \label{app:quad}

The quadratic constraint on the embedding tensor is required in order for the algebra of the gauge group to close
\begin{equation}
 \left[ X_{\mathcal{M}}, X_{\mathcal{N}} \right] = - X_{ \mathcal{M} \mathcal{N}}{}^{\mathcal{P}} X_{\mathcal{P}},
\end{equation}
 or equivalently, 
\begin{equation}
 X_{\cM \cQ}{}^{\cR} X_{\cN \cR}{}^{\cP} - X_{\cN \cQ}{}^{\cR} X_{\cM \cR}{}^{\cP} = - X_{\cM \cN}{}^{\cR} X_{\cR \cQ}{}^{\cP}.
\label{qcons}
\end{equation}
Note that this constraint is highly non-trivial even to the extent that the left hand side of the above equations is manifestly antisymmetric under the interchange of indices $\cM$ and $\cN$, whereas 
$$X_{ \mathcal{M} \mathcal{N}}{}^{\mathcal{P}} $$
is not in general antisymmetric under such an operation. 
We can therefore split this object into two tensors, {\it viz.}
\begin{equation}
 X_{ \mathcal{M} \mathcal{N}}{}^{\mathcal{P}} = X_{ [\mathcal{M} \mathcal{N}]}{}^{\mathcal{P}} + Z_{\mathcal{M} \mathcal{N}}{}^{\mathcal{P}},
\end{equation}
where the components of $X_{ \mathcal{M} \mathcal{N}}{}^{\mathcal{P}}$ in a GL(7) decomposition is given in \eqref{X2} and 
$$ Z_{\mathcal{M} \mathcal{N}}{}^{\mathcal{P}} \equiv
 X_{(\mathcal{M} \mathcal{N})}{}^{\mathcal{P}}.$$
In \eqref{X2} we had already derived all the components of $X_{\mathcal{M} \mathcal{N}}{}^{\mathcal{P}}$ from the generalised vielbein postulates, so we can now explicitly exhibit
the non-zero components of the symmetric tensor $Z_{\cM\cN}{}^\cP$ as
\begin{align}
 Z_{m8}{\,}^{p8}{\,}_{r8} &= Z^{p8}{\,}_{m8}{\,}_{r8} = -\frac{1}{4} f^{p}{}_{mr}, 
\hspace{27.1mm} Z_{m8}{\,}^{pq}{\,}_{r8} = Z^{pq}{\,}_{m8}{\,}_{r8} =  \frac{\sqrt{2}}{2} \delta^{pq}_{mr} \fr, \notag \\[5pt]
Z_{m8}{\,}^{pq}{\,}_{rs} &=  Z^{pq}{\,}_{m8}{\,}_{rs} = - \frac{1}{2} \delta^{[p}_{m} f^{q]}{}_{rs}, 
\hspace{23.3mm} Z_{m8}{\,}_{rs}{\,}^{pq} = Z_{rs}{\,}_{m8}{\,}^{pq} = -\frac{3}{2} \delta^{[p}_{[r} f^{q]}{}_{mn]}, \notag \\[5pt]
 Z_{mn}{\,}^{pq}{\,}_{r8} &= \;\, Z^{pq}{\,}_{mn}{\,}_{r8} \; = - \delta^{[p}_{[m} f^{q]}{}_{n]r}, 
\hspace{20.9mm} Z_{m8}{\,}_{rs}{\,}_{p8} = \;\, Z_{rs}{\,}_{m8}{\,}_{p8} \;\, =   \frac{\sqrt{2}}{4} g_{mprs}, \notag \\[5pt]
& \hspace{20mm} Z^{mn}{\,}_{p8}{\,}^{rs} = Z_{p8}{\,}^{mn}{\,}^{rs} = - \frac{\sqrt{2}}{16} \delta^{[r}_{[p} \eta^{s]mntuvw} g_{tu]vw}, \notag \\[5pt]
& \hspace{11mm} Z^{mn}{\,}^{rs}{\,}_{p8} =  Z^{rs}{\,}^{mn}{\,}_{p8} = \frac{\sqrt{2}}{48} (\delta^{[m}_{p} \eta^{n]rstuvw} + \delta^{[r}_{p} \eta^{s]mntuvw} ) g_{mtuv}, \notag \\[5pt] 
& \hspace{15mm} Z^{mn}{\,}^{pq}{\,}^{rs} = Z^{pq}{\,}^{mn}{\,}^{rs} = \frac{1}{4} (\eta^{pqrstu[m} f^{n]}{}_{tu} + \eta^{mnrstu[p} f^{q]}{}_{tu}) .  
  \label{Z}
\end{align}
The contraction given on the right hand side of equation \eqref{qcons} is indeed 
symmetric under the interchange of $\cM$ and $\cN$ \cite{dWSTmax4}. 

The components of $X_{ \mathcal{M} \mathcal{N}}{}^{\mathcal{P}}$ as derived from the generalised vielbein postulates, \eqref{X2}, satisfy the linear constraint since they can be put into a form compatible with the general solution of the linear constraint \eqref{X} (see section \ref{sec:embedding}). However, the quadratic constraint is not necessarily satisfied by the general solution \eqref{X} and equation \eqref{qcons} must be considered for the particular solution given by equations \eqref{X2}.

The components of $X$, given in \eqref{X2}, satisfy
\begin{equation}
 X_{\cM}{}^{\tP\tQ}{}_{\tR\tS} = - X_{\cM}{}_{\tR\tS}{}^{\tP\tQ}, \qquad  X_{\cM}{}^{\tP\tQ}{}^{\tR\tS} = X_{\cM}{}^{\tR\tS}{}^{\tP\tQ}, \qquad X_{\cM}{}_{\tP\tQ}{}_{\tR\tS} = X_{\cM}{}_{\tR\tS}{}_{\tP\tQ}.
\label{Xsyms}
\end{equation}

We will verify equation \eqref{qcons} for each component in turn:
\begin{enumerate}
 \item \begin{equation}
 X_{\tM\tN \tP\tQ}{}^{\cR} X_{\tT\tU \cR}{}_{\tV\tW} - X_{\tT\tU \tP\tQ}{}^{\cR} X_{\tM\tN \cR}{}_{\tV\tW} = - X_{\tM\tN \tT\tU}{}^{\cR} X_{\cR \tP\tQ}{}_{\tV\tW}.
\label{qcons1}
\end{equation} \newline
The only components for which both sides of the above equation are non-trivial are
$$ (\tM\tN, \tP\tQ, \tT\tU, \tV\tW) = (m8, p8, t8, vw) \quad \textup{or } \quad (m8, pq, t8, v8).$$
The latter case above is equivalent to the former, since from equation \eqref{Xsyms} both sides of equation \eqref{qcons1} are symmetric under the interchange of $\tP\tQ$ and $\tV\tW$. Therefore, we only need to consider 
\begin{align*}
 & X_{m8 \, p8}{\,}^{\cR} X_{t8 \, \cR}{\,}_{vw} - X_{t8 \, p8}{\,}^{\cR}  X_{m8 \, \cR}{\,}_{vw} + X_{m8 \, t8}{\,}^{\cR} X_{\cR \, p8}{\,}_{vw} \\[8pt]
  =& \left[ 2 X_{m8 \, p8}{\,}^{r8} X_{t8 \, r8}{\,}_{vw} +  X_{m8 \, p8}{\,}_{rs} X_{t8 \,}{}^{rs}{\,}_{vw} -  (m \longleftrightarrow t) \right] + X_{m8 \, t8}{\,}^{r8} X_{r8 \, p8}{\,}_{vw} + X_{m8 \, t8}{\,}_{rs} X^{rs}{}_{\, p8}{\,}_{vw} ,  \\[5pt]
 &=- \left[ \frac{\sqrt{2}}{2} f^{r}{}_{mp} g_{trvw} + \sqrt{2} g_{mp[v|s} f^{s}{}_{|w] t}  -  (m \longleftrightarrow t) \right] - \frac{\sqrt{2}}{2} f^{r}{}_{tm} g_{pvwr} - \frac{3\sqrt{2}}{2} f^{r}{}_{[vw} g_{p]tmr}, \\
&= - \frac{3 \sqrt{2}}{2} f^{r}{}_{m[p} g_{vw]tr} - \frac{\sqrt{2}}{4} f^{r}{}_{tm} g_{pvwr} - \frac{3\sqrt{2}}{4} f^{r}{}_{[vw} g_{p]tmr} -  (m \longleftrightarrow t), \\[5pt]
& = - 5 \sqrt{2} f^{r}{}_{[tm} g_{pvw]r}, 
\end{align*}
which vanishes by equation \eqref{fgjac}.

\item
\begin{equation}
  X_{\tM\tN}{}_{\tP\tQ}{}^{\cR} X_{\tT\tU \cR}{}^{\tV\tW} - X_{\tT\tU}{}_{\tP\tQ}{}^{\cR} X_{\tM\tN \cR}{}^{\tV\tW} = - X_{\tM\tN \tT\tU}{}^{\cR} X_{\cR}{}_{\tP\tQ}{}^{\tV\tW}.
\label{qcons2}
\end{equation} \newline
The components of the above equation where both sides of the equation are non-trivial are given by
\begin{equation}
 (\tM\tN, \tP\tQ, \tT\tU, \tV\tW) = 
\begin{cases}
 (m8, p8, t8, v8) \\
(m8, p8, tu, v8) \\
(m8, p8, t8, vw) \\
(mn, p8, t8, vw)
\end{cases}. \label{qcons2cases}
\end{equation}
In the first case, we have 
\begin{align*}
 X_{m8}{\,}_{p8}{\,}^{\cR} X_{t8 \, \cR}{\,}^{v8} - X_{t8}{\,}_{p8}{\,}^{\cR} X_{m8 \, \cR}{\,}^{v8} + X_{m8 \, t8}{\,}^{\cR} X_{\cR}{\,}_{p8}{\,}^{v8} &= - f^{s}{}_{p[t|} f^{v}_{s|m]} + \frac{1}{2} f^{s}{}_{tm} f^{v}{}_{ps}, \\
&= \frac{3}{2}  f^{s}{}_{[tm} f^{v}{}_{p]s},
\end{align*}
which vanishes by equation \eqref{ffjac}. Similarly, the second case also vanishes by equation \eqref{ffjac}. 

Consider the third case in \eqref{qcons2cases}, 
\begin{align*}
& X_{m8}{\,}_{p8}{\,}^{\cR} X_{t8 \, \cR}{\,}^{vw} - X_{t8}{\,}_{p8}{\,}^{\cR} X_{m8 \, \cR}{\,}^{vw} + X_{m8 \, t8}{\,}^{\cR} X_{\cR}{\,}_{p8}{\,}^{vw} \\[5pt]
=& -\frac{1}{6} \eta^{vw r_1 \dots r_5} g_{[m|r_1 r_2 r_3} g_{|t] r_4 r_5 p} + \frac{1}{24} \delta^{[v}_{p} \eta^{w] r_1 \dots r_6} g_{mtr_1 r_2} g_{r_3 \dots r_6} , \\[5pt]
=&- \frac{1}{6} \eta^{vw r_1 \dots r_5} g_{[m|r_1 r_2 r_3} g_{|t] r_4 r_5 p} + \frac{1}{6} \delta^{[v}_{p} \eta^{w r_1 \dots r_6]} g_{mtr_1 r_2} g_{r_3 \dots r_6} - \frac{1}{8} \eta^{v w r_1 \dots r_5} g_{mt [p r_1} g_{r_2 \dots r_5]} , \\[3pt]
=&\frac{1}{6} \delta^{[v}_{p} \eta^{w r_1 \dots r_6]} g_{mtr_1 r_2} g_{r_3 \dots r_6} - \frac{7}{24} \eta^{vw r_1 \dots r_5} g_{[mt p r_1} g_{r_2 \dots r_5]}.
\end{align*}
Both of the terms above vanish because they contain antisymmetrisations over 8 indices. Moreover, it is simple to show that equation \eqref{qcons2} is satisfied for the fourth case, as in this case both sides of equation \eqref{qcons} are equal to
$$\delta^{[v}_{p} f^{w]}{}_{ts} f^{s}{}_{mn}.$$

\item \begin{equation}
 X_{\tM\tN}{}^{\tP\tQ}{}^{\cR} X_{\tT\tU \cR}{}_{\tV\tW} - X_{\tT\tU}{}^{\tP\tQ}{}^{\cR} X_{\tM\tN \cR}{}_{\tV\tW} = - X_{\tM\tN \tT\tU}{}^{\cR} X_{\cR}{}^{\tP\tQ}{}_{\tV\tW}.
\label{qcons3}
\end{equation} \newline
Using the identities given in \eqref{Xsyms}, the above equation reduces to  
 \begin{equation} 
  X_{\tM\tN}{}^{\cR}{}^{\tP\tQ} X_{\tT\tU}{}_{\tV\tW}{}_{\cR} - X_{\tT\tU}{}^{\cR}{}^{\tP\tQ} X_{\tM\tN}{}_{\tV\tW}{}_{\cR} =  X_{\tM\tN \tT\tU}{}^{\cR} X_{\cR}{}_{\tV\tW}{}^{\tP\tQ},
 \end{equation}
which is equivalent to equation \eqref{qcons2}.

\item 
\begin{equation}
 X_{\tM\tN}{}^{\tP\tQ}{}^{\cR} X_{\tT\tU \cR}{}^{\tV\tW} - X_{\tT\tU}{}^{\tP\tQ}{}^{\cR} X_{\tM\tN \cR}{}^{\tV\tW} = - X_{\tM\tN \tT\tU}{}^{\cR} X_{\cR}{}^{\tP\tQ}{}^{\tV\tW}
\label{qcons4}
\end{equation} \newline
There is only one component of equation \eqref{qcons4} for which both sides of the above equation are non-vanishing:
\begin{align*}
  & X_{m8}{\,}^{pq}{\,}^{\cR} X_{t8 \, \cR}{\,}^{vw} - X_{t8}{\,}^{pq}{\,}^{\cR} X_{m8 \, \cR}{\,}^{vw} + X_{m8 \,t8}{\,}^{\cR} X_{\cR}{\,}^{pq}{\,}^{vw} \\[5pt]
=& - \frac{\sqrt{2}}{3} \eta^{pqr_1 \dots r_4 [v} g_{[m|r_1 r_2 r_3} f^{w]}{}_{r_4|t]} - \frac{\sqrt{2}}{3} \eta^{vwr_1 \dots r_4 [p} g_{[m|r_1 r_2 r_3} f^{q]}{}_{r_4|t]} \\[3pt]
& \hspace{30mm} + \frac{\sqrt{2}}{4} \eta^{pqvwu_1 u_2 u_3} f^{s}{}_{u_1 u_2} g_{mtu_3s}  + \frac{\sqrt{2}}{12} \eta^{pqvw u_1 u_2 u_3} f^{s}{}_{mt} g_{u_1 u_2 u_3 s} , \\[5pt]
=& - \frac{2 \sqrt{2}}{3} \eta^{r_1 \dots r_4 [pqv} g_{[m|r_1 r_2 r_3} f^{w]}{}_{r_4|t]} + \frac{5 \sqrt{2}}{6} \eta^{pqvw u_1 u_2 u_3} f^{s}{}_{[mt} g_{u_1 u_2 u_3] s} + \frac{\sqrt{2}}{2} \eta^{pqvw u_1 u_2 u_3} f^{s}{}_{u_1[t} g_{m] u_2 u_3 s}, \\
=& - \frac{4 \sqrt{2}}{3} \eta^{[r_1 \dots r_4 pqv} g_{[m|r_1 r_2 r_3} f^{w]}{}_{r_4|t]} + \frac{\sqrt{2}}{6} \eta^{pqvwr_1 \dots r_3} g_{[m|r_1 r_2 r_3} f^{s}{}_{s|t]} + \frac{5 \sqrt{2}}{6} \eta^{pqvw u_1 u_2 u_3} f^{s}{}_{[mt} g_{u_1 u_2 u_3] s},
\end{align*}
which vanishes by unimodularity, \eqref{funi}, and equation \eqref{fgjac}. 

\item 
\begin{equation}
 X_{\tM\tN}{}_{\tP\tQ}{}^{\cR} X^{\tT\tU}{}_{\cR}{}_{\tV\tW} - X^{\tT\tU}{}_{\tP\tQ}{}^{\cR} X_{\tM\tN \cR}{}_{\tV\tW} = - X_{\tM\tN}{}^{\tT\tU}{}^{\cR} X_{\cR}{}_{\tP\tQ}{}_{\tV\tW}
\label{qcons5}
\end{equation} \newline
The only non-trivial components to consider in this case are
\begin{equation}
 (\tM\tN, \tP\tQ, \tT\tU, \tV\tW) = (m8, p8, tu, vw) \quad \textup{or} \quad (m8, pq, tu, v8)
\end{equation}
Both cases reduce to the same equation, hence we only consider the first case:
\begin{align*}
  X_{m8}{\,}_{p8}{\,}^{\cR}  X^{tu}{\,}_{\cR}{\,}_{vw} - X^{tu}{\,}_{p8}{\,}^{\cR} & X_{m8 \, \cR}{\,}_{vw}  + X_{m8}{\,}^{tu}{\,}^{\cR} X_{\cR}{\,}_{p8}{\,}_{vw} \\[3pt] 
&= 6 \delta^{r}_{[v} f^{s}{}_{w]m} \delta^{[t}_{[p} f^{u]}{}_{rs]} + 3 f^{r}{}_{pm} \delta^{[t}_{[r} f^{v]}{}_{vw]} - 6 \delta^{[r}_{[p} f^{s]}{}_{vw]} \delta^{[t}_{r} f^{u]}{}_{sm}, \\[3pt]
&= 3  \delta^{[t|}_{v} f^{s}{}_{[pm} f^{|u]}{}_{w]s} - 3  \delta^{[t|}_{w} f^{s}{}_{[pm} f^{|u]}{}_{v]s} + 3  \delta^{[t|}_{p} f^{s}{}_{[vw} f^{|u]}{}_{m]s},
\end{align*}
which vanishes by equation \eqref{ffjac}.

\item 
\begin{equation}
 X_{\tM\tN}{}_{\tP\tQ}{}^{\cR} X^{\tT\tU}{}_{\cR}{}^{\tV\tW} - X^{\tT\tU}{}_{\tP\tQ}{}^{\cR} X_{\tM\tN \cR}{}^{\tV\tW} = - X_{\tM\tN}{}^{\tT\tU}{}^{\cR} X_{\cR}{}_{\tP\tQ}{}^{\tV\tW}
\label{qcons6}
\end{equation} \newline
It is straightforward to see that all terms in the above equation vanish trivially unless
\begin{equation}
 (\tM\tN, \tP\tQ, \tT\tU, \tV\tW) = (m8, p8, tu, vw).
\end{equation}
In this case,
\begin{align*}
 & X_{m8}{\,}_{p8}{\,}^{\cR} X^{tu}{\,}_{\cR}{\,}^{vw} - X^{tu}{\,}_{p8}{\,}^{\cR} X_{m8 \, \cR}{\,}^{vw} + X_{m8}{\,}^{tu}{\,}^{\cR} X_{\cR}{\,}_{p8}{\,}^{vw} \\[5pt]
=& - \frac{\sqrt{2}}{24} f^{[v}{}_{mp} \eta^{w] tu s_1 \dots s_4} g_{s_1 \dots s_4} - \frac{\sqrt{2}}{4} f^{[t}{}_{s_1 s_2} \eta^{u] vw s_1 \dots s_4} g_{mp s_3 s_4} - \frac{\sqrt{2}}{12} \delta^{[v}_{[p} f^{w]}{}_{s] m} \eta^{s tu q_1 \dots q_4} g_{q_1 \dots q_4} \\[5pt]
& - \frac{\sqrt{2}}{4} \delta^{[t}_{[p} f^{u]}{}_{rs]} \eta^{rs vw q_1 \dots q_3} g_{m q_1 \dots q_3}  - \frac{\sqrt{2}}{12} \delta^{[t}_{r} f^{u]}{}_{s m} \delta^{[v}_{p} \eta^{w] rs q_1 \dots q_4} g_{q_1 \dots q_4}
+ \frac{\sqrt{2}}{12} \delta^{[v}_{p} f^{w]}{}_{rs} \eta^{tu rs q_1 \dots q_3} g_{m q_1 \dots q_3}.
\end{align*}
Using Schouten identities, the first, third and fifth terms in the expression on the right hand side reduce to
\begin{equation}
 \frac{\sqrt{2}}{6} \delta^{[v}_{p} \eta^{w] tu r_1 \dots r_4} f^{s}{}_{mr_1} g_{r_2 \dots r_4 s}  
\end{equation}
 and similarly the second and fourth term simplify to
\begin{equation} 
- \frac{\sqrt{2}}{6} f^{[t}{}_{r_1 r_2} \eta^{u] [v| r_1 \dots r_5 } \delta^{|w]}_{p} g_{m r_3 \dots r_5}.
\end{equation}
Therefore, 
\begin{align*}
 & X_{m8}{\,}_{p8}{\,}^{\cR} X^{tu}{\,}_{\cR}{\,}^{vw} - X^{tu}{\,}_{p8}{\,}^{\cR} X_{m8 \, \cR}{\,}^{vw} + X_{m8}{\,}^{tu}{\,}^{\cR} X_{\cR}{\,}_{p8}{\,}^{vw} \\[5pt]
=& \frac{\sqrt{2}}{6} \delta^{[v}_{p} \eta^{w] tu r_1 \dots r_4} f^{s}{}_{mr_1} g_{r_2 \dots r_4 s} - \frac{\sqrt{2}}{6} f^{[t}{}_{r_1 r_2} \eta^{u] [v| r_1 \dots r_5 } \delta^{|w]}_{p} g_{m r_3 \dots r_5}
+ \frac{\sqrt{2}}{12} \delta^{[v}_{p} f^{w]}{}_{rs} \eta^{tu rs q_1 \dots q_3} g_{m q_1 \dots q_3}, \\[5pt]
=& \frac{5 \sqrt{2}}{24} \delta^{[v}_{p} \eta^{w] tu r_1 \dots r_4} f^{s}{}_{[mr_1} g_{r_2 \dots r_4] s} + \frac{\sqrt{2}}{6} \delta^{[v}_{p} \eta^{w] tu r_1 \dots r_4} f^{s}{}_{s r_1} g_{m r_2 \dots r_4},
\end{align*}
where we have again used Schouten identities. It is now clear that equation \eqref{qcons6} holds as a result of equations \eqref{funi} and \eqref{fgjac}.

\item 
\begin{equation}
 X_{\tM\tN}{}^{\tP\tQ}{}^{\cR} X^{\tT\tU}{}_{\cR}{}_{\tV\tW} - X^{\tT\tU}{}^{\tP\tQ}{}^{\cR} X_{\tM\tN \cR}{}_{\tV\tW} = - X_{\tM\tN}{}^{\tT\tU}{}^{\cR} X_{\cR}{}^{\tP\tQ}{}_{\tV\tW}.
\label{qcons7}
\end{equation} \newline
Using the relations in \eqref{Xsyms}, this equation is equivalent to equation \eqref{qcons6}, which we have already verified. 

\item 
\begin{equation}
 X_{\tM\tN}{}^{\tP\tQ}{}^{\cR} X^{\tT\tU}{}_{\cR}{}^{\tV\tW} - X^{\tT\tU}{}^{\tP\tQ}{}^{\cR} X_{\tM\tN \cR}{}^{\tV\tW} = - X_{\tM\tN}{}^{\tT\tU}{}^{\cR} X_{\cR}{}^{\tP\tQ}{}^{\tV\tW}.
\label{qcons8}
\end{equation} \newline
The only non-trivial equation to consider in this case is 
\begin{align*}
  X_{m8}{\,}^{pq}{\,}^{\cR} X^{tu}{\,}_{\cR}{\,}^{vw} - X^{tu}{\,}^{pq}{\,}^{\cR} X_{m8 \, \cR}{\,}^{vw} + X_{m8}{\,}^{tu}{\,}^{\cR} X_{\cR}{\,}^{pq}{\,}^{vw} = \frac{3}{2} \eta^{pqvw s_1 s_2 [t} f^{u]}{}_{r [m} f^{r}{}_{s_1 s_2]},
\end{align*}
where we have used Schouten identities. Therefore, equation \eqref{qcons8} is satisfied.

\item 
\begin{equation}
 X^{\tM\tN}{}_{\cQ}{}^{\cR} X_{\tT\tU \cR}{}^{\cP} - X_{\tT\tU \cQ}{}^{\cR} X^{\tM\tN}{}_{\cR}{}^{\cP} = - X^{\tM\tN}{}_{\tT\tU}{}^{\cR} X_{\cR \cQ}{}^{\cP}.
\label{qcons9}
\end{equation} \newline
Note that the left hand side of this equation is of the same form as the left hand side of cases 5--8. Therefore, it remains to show that 
\begin{equation}
 - X^{\tM\tN}{}_{\tT\tU}{}^{\cR} X_{\cR \cQ}{}^{\cP} = X_{\tT\tU}{}^{\tM\tN}{}^{\cR} X_{\cR \cQ}{}^{\cP}.
\end{equation}
This can be simply verified using Schouten identities and equations \eqref{funi}, \eqref{ffjac} and \eqref{fgjac} for all components.

\item 
\begin{equation}
 X^{\tM\tN}{}_{\tP\tQ}{}^{\cR} X^{\tT\tU}{}_{\cR}{}_{\tV\tW} - X^{\tT\tU}{}_{\tP\tQ}{}^{\cR} X^{\tM\tN}{}_{\cR}{}_{\tV\tW} = - X^{\tM\tN}{}^{\tT\tU}{}^{\cR} X_{\cR}{}_{\tP\tQ}{}_{\tV\tW}.
\label{qcons10}
\end{equation} \newline
This equation is trivially satisfied. 

\item 
\begin{equation}
 X^{\tM\tN}{}_{\tP\tQ}{}^{\cR} X^{\tT\tU}{}_{\cR}{}^{\tV\tW} - X^{\tT\tU}{}_{\tP\tQ}{}^{\cR} X^{\tM\tN}{}_{\cR}{}^{\tV\tW} = - X^{\tM\tN}{}^{\tT\tU}{}^{\cR} X_{\cR}{}_{\tP\tQ}{}^{\tV\tW}.
\label{qcons11}
\end{equation} \newline
The only non-trivial components to consider is
\begin{align*}
  & X^{mn}{\,}_{p8}{\,}^{\cR} X^{tu}{\,}_{\cR}{\,}^{vw} - X^{tu}{\,}_{p8}{\,}^{\cR} X^{mn}{\,}_{\cR}{\,}^{vw} + X^{mn}{\,}^{tu}{\,}^{\cR} X_{\cR}{\,}_{p8}{\,}^{vw} \\[5pt]
=&  \frac{3}{2} \eta^{vwrsq_1 q_2 [m} f^{n]}{}_{q_1 q_2} \delta^{[t}_{[r} f^{u]}{}_{sp]} - \frac{3}{2} \eta^{vwrsq_1 q_2 [t} f^{u]}{}_{q_1 q_2} \delta^{[m}_{[r} f^{n]}{}_{sp]} - \frac{1}{2} \eta^{tursq_1 q_2[m} f^{n]}{}_{q_1 q_2} \delta^{[v}_{p} f^{w]}{}_{rs}, \\[5pt]
=& \frac{1}{2} \delta^{[v}_{p} \eta^{w]m [t|r_1 \dots r_4} f^{n}{}_{r_1 r_2} f^{|u]}{}_{r_3 r_4} - \frac{1}{2} \delta^{[v}_{p} \eta^{w] n [t|r_1 \dots r_4} f^{m}{}_{r_1 r_2} f^{|u]}{}_{r_3 r_4} - \frac{1}{2} \eta^{tursq_1 q_2[m} f^{n]}{}_{q_1 q_2} \delta^{[v}_{p} f^{w]}{}_{rs},
\end{align*}
where in the second equality we have used Schouten identities to simplify the first two terms on the second line. Further use of Schouten identities gives 
\begin{align*}
 X^{mn}{\,}_{p8}{\,}^{\cR} X^{tu}{\,}_{\cR}{\,}^{vw} - &X^{tu}{\,}_{p8}{\,}^{\cR} X^{mn}{\,}_{\cR}{\,}^{vw} + X^{mn}{\,}^{tu}{\,}^{\cR} X_{\cR}{\,}_{p8}{\,}^{vw} \\[5pt]
=& \frac{1}{2} \delta^{[v}_{p} \eta^{w] tu r_1 \dots r_4} f^{[m}{}_{r_1 r_2} f^{n]}{}_{r_3 r_4} + 2 \delta^{[v}_{p} \eta^{w]tu r_1 r_2 r_3 [m} f^{n]}{}_{[r_1 r_2} f^{r_4}{}_{r_3 r_4]}.
\end{align*}
The first term vanishes as a consequence of the fact that 
$$f^{[m}{}_{[r_1 r_2} f^{n]}{}_{r_3 r_4]}$$
is antisymmetric under the interchange of $m$ and $n$, but symmetric under the interchange of pairs $[r_1 r_2]$ and $[r_3 r_4]$. Furthermore, the second term vanishes either by the unimodularity property \eqref{funi} or the Jacobi identity \eqref{ffjac}. Hence equation \eqref{qcons11} is satisfied.

\item 
\begin{equation}
 X^{\tM\tN}{}^{\tP\tQ}{}^{\cR} X^{\tT\tU}{}_{\cR}{}_{\tV\tW} - X^{\tT\tU}{}^{\tP\tQ}{}^{\cR} X^{\tM\tN}{}_{\cR}{}_{\tV\tW} = - X^{\tM\tN}{}^{\tT\tU}{}^{\cR} X_{\cR}{}^{\tP\tQ}{}_{\tV\tW}.
\label{qcons12}
\end{equation} \newline
Using equations \eqref{Xsyms}, this case is equivalent to case 11, which we have already verified.

\item 
\begin{equation}
 X^{\tM\tN}{}^{\tP\tQ}{}^{\cR} X^{\tT\tU}{}_{\cR}{}^{\tV\tW} - X^{\tT\tU}{}^{\tP\tQ}{}^{\cR} X^{\tM\tN}{}_{\cR}{}^{\tV\tW} = - X^{\tM\tN}{}^{\tT\tU}{}^{\cR} X_{\cR}{}^{\tP\tQ}{}^{\tV\tW}.
\label{qcons13}
\end{equation} \newline
The above equation is trivially satisfied.
\end{enumerate}

\newpage

\bibliography{ss}

\providecommand{\href}[2]{#2}\begingroup\raggedright\begin{thebibliography}{10}

\bibitem{GGN13}
H.~Godazgar, M.~Godazgar, and H.~Nicolai, ``{Generalised geometry from the
  ground up},''
\href{http://arxiv.org/abs/1307.8295}{{\tt arXiv:1307.8295 [hep-th]}}.

\bibitem{CJS}
E.~Cremmer, B.~Julia, and J.~Scherk, ``{Supergravity theory in
  eleven-dimensions},''
  \href{http://dx.doi.org/10.1016/0370-2693(78)90894-8}{{\em Phys.Lett.} {\bf
  B76} (1978)  409--412}.

\bibitem{cremmerjulia}
E.~Cremmer and B.~Julia, ``{The N=8 supergravity theory. 1. The Lagrangian},''
  \href{http://dx.doi.org/10.1016/0370-2693(78)90303-9}{{\em Phys.Lett.} {\bf
  B80} (1978)  48}.

\bibitem{dWNsu8}
B.~de~Wit and H.~Nicolai, ``{d} = 11 supergravity with local {SU}(8)
  invariance,'' \href{http://dx.doi.org/10.1016/0550-3213(86)90290-7}{{\em
  Nucl.Phys.} {\bf B274} (1986)  363}.

\bibitem{dWN13}
B.~de~Wit and H.~Nicolai, ``{Deformations of gauged SO(8) supergravity and
  supergravity in eleven dimensions},''
  \href{http://dx.doi.org/10.1007/JHEP05(2013)077}{{\em JHEP} {\bf 1305} (2013)
   077},
\href{http://arxiv.org/abs/1302.6219}{{\tt arXiv:1302.6219 [hep-th]}}.

\bibitem{Nso16}
H.~Nicolai, ``{D} = 11 supergravity with local {SO(16)} invariance,''
\href{http://dx.doi.org/10.1016/0370-2693(87)91102-6}{{\em Phys.Lett.} {\bf
  B187} (1987)  316}.

\bibitem{KNS}
K.~Koepsell, H.~Nicolai, and H.~Samtleben, ``{An exceptional geometry for D =
  11 supergravity?},''
  \href{http://dx.doi.org/10.1088/0264-9381/17/18/308}{{\em Class.Quant.Grav.}
  {\bf 17} (2000)  3689--3702}, \href{http://arxiv.org/abs/hep-th/0006034}{{\tt
  arXiv:hep-th/0006034 [hep-th]}}.

\bibitem{btreview}
D.~S. Berman and D.~C. Thompson, ``{Duality symmetric string and M-theory},''
\href{http://arxiv.org/abs/1306.2643}{{\tt arXiv:1306.2643 [hep-th]}}.

\bibitem{Hohm:2013vpa}
O.~Hohm and H.~Samtleben, ``{Exceptional Field Theory I: $E_{6(6)}$ covariant
  Form of M-Theory and Type IIB},''
\href{http://arxiv.org/abs/1312.0614}{{\tt arXiv:1312.0614 [hep-th]}}.

\bibitem{dWNW}
B.~de~Wit, H.~Nicolai, and N.~P. Warner, ``The embedding of gauged {N}=8
  supergravity into {d = 11} supergravity,''
\href{http://dx.doi.org/10.1016/0550-3213(85)90128-2}{{\em Nucl.Phys.} {\bf
  B255} (1985)  29}.

\bibitem{dWNconsis}
B.~de~Wit and H.~Nicolai, ``{The consistency of the S**7 truncation in D=11
  supergravity},''
\href{http://dx.doi.org/10.1016/0550-3213(87)90253-7}{{\em Nucl.Phys.} {\bf
  B281} (1987)  211}.

\bibitem{NP}
H.~Nicolai and K.~Pilch, ``{Consistent truncation of d = 11 supergravity on
  AdS$_4 \times S^7$},'' \href{http://dx.doi.org/10.1007/JHEP03(2012)099}{{\em
  JHEP} {\bf 1203} (2012)  099},
\href{http://arxiv.org/abs/1112.6131}{{\tt arXiv:1112.6131 [hep-th]}}.

\bibitem{duffpope}
M.~J. Duff and C.~N. Pope, ``Kaluza-{K}lein supergravity and the seven
  sphere,'' in {\em Supersymmetry and supergravity 82}, S.~Ferrara, J.~G.
  Taylor, and P.~van Nieuwenhuizen, eds.
\newblock World Scientific, 1983.

\bibitem{GGN}
H.~Godazgar, M.~Godazgar, and H.~Nicolai, ``{Testing the non-linear flux ansatz
  for maximal supergravity},''
  \href{http://dx.doi.org/10.1103/PhysRevD.87.085038}{{\em Phys.Rev.} {\bf D87}
  (2013)  085038},
\href{http://arxiv.org/abs/1303.1013}{{\tt arXiv:1303.1013 [hep-th]}}.

\bibitem{KKdual}
H.~Godazgar, M.~Godazgar, and H.~Nicolai, ``{Non-linear Kaluza-Klein theory for
  dual fields},''
\href{http://arxiv.org/abs/1309.0266}{{\tt arXiv:1309.0266 [hep-th]}}.

\bibitem{dWSTlag}
B.~de~Wit, H.~Samtleben, and M.~Trigiante, ``{On Lagrangians and gaugings of
  maximal supergravities},''
  \href{http://dx.doi.org/10.1016/S0550-3213(03)00059-2}{{\em Nucl.Phys.} {\bf
  B655} (2003)  93--126},
\href{http://arxiv.org/abs/hep-th/0212239}{{\tt arXiv:hep-th/0212239
  [hep-th]}}.

\bibitem{dWSTgauge}
B.~de~Wit, H.~Samtleben, and M.~Trigiante, ``{Gauging maximal
  supergravities},'' \href{http://dx.doi.org/10.1002/prop.200410135}{{\em
  Fortsch.Phys.} {\bf 52} (2004)  489--496},
\href{http://arxiv.org/abs/hep-th/0311225}{{\tt arXiv:hep-th/0311225
  [hep-th]}}.

\bibitem{dWSTmax4}
B.~de~Wit, H.~Samtleben, and M.~Trigiante, ``{The maximal D=4
  supergravities},''
  \href{http://dx.doi.org/10.1088/1126-6708/2007/06/049}{{\em JHEP} {\bf 0706}
  (2007)  049},
\href{http://arxiv.org/abs/0705.2101}{{\tt arXiv:0705.2101 [hep-th]}}.

\bibitem{NSmaximal3}
H.~Nicolai and H.~Samtleben, ``{Maximal gauged supergravity in
  three-dimensions},''
  \href{http://dx.doi.org/10.1103/PhysRevLett.86.1686}{{\em Phys.Rev.Lett.}
  {\bf 86} (2001)  1686--1689},
\href{http://arxiv.org/abs/hep-th/0010076}{{\tt arXiv:hep-th/0010076
  [hep-th]}}.

\bibitem{NScomgauge3}
H.~Nicolai and H.~Samtleben, ``{Compact and noncompact gauged maximal
  supergravities in three-dimensions},'' {\em JHEP} {\bf 0104} (2001)  022,
\href{http://arxiv.org/abs/hep-th/0103032}{{\tt arXiv:hep-th/0103032
  [hep-th]}}.

\bibitem{dWNn8}
B.~de~Wit and H.~Nicolai, ``{N}=8 supergravity,''
\href{http://dx.doi.org/10.1016/0550-3213(82)90120-1}{{\em Nucl.Phys.} {\bf
  B208} (1982)  323}.

\bibitem{scherkschwarz}
J.~Scherk and J.~H. Schwarz, ``{How to get masses from extra dimensions},''
\href{http://dx.doi.org/10.1016/0550-3213(79)90592-3}{{\em Nucl.Phys.} {\bf
  B153} (1979)  61--88}.

\bibitem{dewitt}
B.~S. DeWitt, ``Dynamical theory of groups and fields,'' in {\em Relativity,
  groups and topology (Les Houches 1963)}, C.~DeWitt and B.~S. Dewitt, eds.
\newblock Gordon and Breach, 1964.

\bibitem{CGP}
M.~Cvetic, G.~Gibbons, H.~Lu, and C.~Pope, ``{Consistent group and coset
  reductions of the bosonic string},''
  \href{http://dx.doi.org/10.1088/0264-9381/20/23/013}{{\em Class.Quant.Grav.}
  {\bf 20} (2003)  5161--5194},
\href{http://arxiv.org/abs/hep-th/0306043}{{\tt arXiv:hep-th/0306043
  [hep-th]}}.

\bibitem{ttf1}
G.~Dall'Agata and S.~Ferrara, ``{Gauged supergravity algebras from twisted tori
  compactifications with fluxes},''
  \href{http://dx.doi.org/10.1016/j.nuclphysb.2005.03.039}{{\em Nucl.Phys.}
  {\bf B717} (2005)  223--245},
\href{http://arxiv.org/abs/hep-th/0502066}{{\tt arXiv:hep-th/0502066
  [hep-th]}}.

\bibitem{ttf2}
L.~Andrianopoli, M.~Lledo, and M.~Trigiante, ``{The Scherk-Schwarz mechanism as
  a flux compactification with internal torsion},''
  \href{http://dx.doi.org/10.1088/1126-6708/2005/05/051}{{\em JHEP} {\bf 0505}
  (2005)  051},
\href{http://arxiv.org/abs/hep-th/0502083}{{\tt arXiv:hep-th/0502083
  [hep-th]}}.

\bibitem{DAF}
G.~Dall'Agata, R.~D'Auria, and S.~Ferrara, ``{Compactifications on twisted tori
  with fluxes and free differential algebras},''
  \href{http://dx.doi.org/10.1016/j.physletb.2005.04.005}{{\em Phys.Lett.} {\bf
  B619} (2005)  149--154},
\href{http://arxiv.org/abs/hep-th/0503122}{{\tt arXiv:hep-th/0503122
  [hep-th]}}.

\bibitem{Mtwistedtorus}
R.~D'Auria, S.~Ferrara, and M.~Trigiante, ``{E(7(7)) symmetry and dual gauge
  algebra of M-theory on a twisted seven-torus},''
  \href{http://dx.doi.org/10.1016/j.nuclphysb.2005.10.020}{{\em Nucl.Phys.}
  {\bf B732} (2006)  389--400},
\href{http://arxiv.org/abs/hep-th/0504108}{{\tt arXiv:hep-th/0504108
  [hep-th]}}.

\bibitem{ttf3}
R.~D'Auria, S.~Ferrara, and M.~Trigiante, ``{Curvatures and potential of
  M-theory in D=4 with fluxes and twist},''
  \href{http://dx.doi.org/10.1088/1126-6708/2005/09/035}{{\em JHEP} {\bf 0509}
  (2005)  035},
\href{http://arxiv.org/abs/hep-th/0507225}{{\tt arXiv:hep-th/0507225
  [hep-th]}}.

\bibitem{dallprez}
G.~Dall'Agata and N.~Prezas, ``{Scherk-Schwarz reduction of M-theory on
  G2-manifolds with fluxes},''
  \href{http://dx.doi.org/10.1088/1126-6708/2005/10/103}{{\em JHEP} {\bf 0510}
  (2005)  103},
\href{http://arxiv.org/abs/hep-th/0509052}{{\tt arXiv:hep-th/0509052
  [hep-th]}}.

\bibitem{Fre}
P.~Fr\'e, ``{M-theory FDA, twisted tori and Chevalley cohomology},''
  \href{http://dx.doi.org/10.1016/j.nuclphysb.2006.02.008}{{\em Nucl.Phys.}
  {\bf B742} (2006)  86--123},
\href{http://arxiv.org/abs/hep-th/0510068}{{\tt arXiv:hep-th/0510068
  [hep-th]}}.

\bibitem{ssdft}
R.~D'Auria, S.~Ferrara, and M.~Trigiante, ``{Supersymmetric completion of
  M-theory 4D-gauge algebra from twisted tori and fluxes},''
  \href{http://dx.doi.org/10.1088/1126-6708/2006/01/081}{{\em JHEP} {\bf 0601}
  (2006)  081},
\href{http://arxiv.org/abs/hep-th/0511158}{{\tt arXiv:hep-th/0511158
  [hep-th]}}.

\bibitem{fretrig}
P.~Fr\'e and M.~Trigiante, ``{Twisted tori and fluxes: A no go theorem for Lie
  groups of weak G(2) holonomy},''
  \href{http://dx.doi.org/10.1016/j.nuclphysb.2006.06.006}{{\em Nucl.Phys.}
  {\bf B751} (2006)  343--375},
\href{http://arxiv.org/abs/hep-th/0603011}{{\tt arXiv:hep-th/0603011
  [hep-th]}}.

\bibitem{ttf4}
C.~Hull and R.~Reid-Edwards, ``{Flux compactifications of M-theory on twisted
  Tori},'' \href{http://dx.doi.org/10.1088/1126-6708/2006/10/086}{{\em JHEP}
  {\bf 0610} (2006)  086},
\href{http://arxiv.org/abs/hep-th/0603094}{{\tt arXiv:hep-th/0603094
  [hep-th]}}.

\bibitem{SamLect}
H.~Samtleben, ``{Lectures on gauged supergravity and flux compactifications},''
  \href{http://dx.doi.org/10.1088/0264-9381/25/21/214002}{{\em
  Class.Quant.Grav.} {\bf 25} (2008)  214002},
\href{http://arxiv.org/abs/0808.4076}{{\tt arXiv:0808.4076 [hep-th]}}.

\bibitem{Grana:2012rr}
M.~Grana and D.~Marques, ``{Gauged double field theory},''
  \href{http://dx.doi.org/10.1007/JHEP04(2012)020}{{\em JHEP} {\bf 1204} (2012)
   020},
\href{http://arxiv.org/abs/1201.2924}{{\tt arXiv:1201.2924 [hep-th]}}.

\bibitem{Berman:2012uy}
D.~S. Berman, E.~T. Musaev, D.~C. Thompson, and D.~C. Thompson, ``{Duality
  invariant M-theory: Gauged supergravities and Scherk-Schwarz reductions},''
  \href{http://dx.doi.org/10.1007/JHEP10(2012)174}{{\em JHEP} {\bf 1210} (2012)
   174},
\href{http://arxiv.org/abs/1208.0020}{{\tt arXiv:1208.0020 [hep-th]}}.

\bibitem{Musaev:2013rq}
E.~T. Musaev, ``{Gauged supergravities in 5 and 6 dimensions from generalised
  Scherk-Schwarz reductions},''
  \href{http://dx.doi.org/10.1007/JHEP05(2013)161}{{\em JHEP} {\bf 1305} (2013)
   161},
\href{http://arxiv.org/abs/1301.0467}{{\tt arXiv:1301.0467 [hep-th]}}.

\bibitem{Aldazabal:2013mya}
G.~Aldazabal, M.~Gra\~na, D.~Marqués, and J.~Rosabal, ``{Extended geometry and
  gauged maximal supergravity},''
  \href{http://dx.doi.org/10.1007/JHEP06(2013)046}{{\em JHEP} {\bf 1306} (2013)
   046},
\href{http://arxiv.org/abs/1302.5419}{{\tt arXiv:1302.5419 [hep-th]}}.

\bibitem{sneddon}
G.~E. {Sneddon},
  \href{http://dx.doi.org/10.1088/0305-4470/9/2/007}{``{Hamiltonian cosmology:
  a further investigation},''{\em Journal of Physics A Mathematical General}
  {\bf 9} (Feb., 1976)  229--238}.

\bibitem{DIT}
G.~Dall'Agata, G.~Inverso, and M.~Trigiante, ``{Evidence for a family of SO(8)
  gauged supergravity theories},''
  \href{http://dx.doi.org/10.1103/PhysRevLett.109.201301}{{\em Phys.Rev.Lett.}
  {\bf 109} (2012)  201301},
\href{http://arxiv.org/abs/1209.0760}{{\tt arXiv:1209.0760 [hep-th]}}.

\bibitem{BHN}
G.~{Bossard}, C.~{Hillmann}, and H.~{Nicolai},
  \href{http://dx.doi.org/10.1007/JHEP12(2010)052}{``{E$_{7(7)}$ symmetry in
  perturbatively quantised $\mathcal{N}$ = 8 supergravity},''{\em Journal of
  High Energy Physics} {\bf 12} (Dec., 2010)  52},
  \href{http://arxiv.org/abs/1007.5472}{{\tt arXiv:1007.5472 [hep-th]}}.

\bibitem{ADFL}
L.~Andrianopoli, R.~D'Auria, S.~Ferrara, and M.~Lledo, ``{Gauging of flat
  groups in four-dimensional supergravity},'' {\em JHEP} {\bf 0207} (2002)
  010,
\href{http://arxiv.org/abs/hep-th/0203206}{{\tt arXiv:hep-th/0203206
  [hep-th]}}.

\bibitem{hillmann}
C.~Hillmann, ``{Generalized E(7(7)) coset dynamics and D=11 supergravity},''
  \href{http://dx.doi.org/10.1088/1126-6708/2009/03/135}{{\em JHEP} {\bf 0903}
  (2009)  135}, \href{http://arxiv.org/abs/0901.1581}{{\tt arXiv:0901.1581
  [hep-th]}}.

\end{thebibliography}\endgroup
\bibliographystyle{utphys}
\end{document}